\documentclass[usenatbib,usegraphicx]{mn2e}
\usepackage{psfig,epsf}

\newcommand{\kms}{\ensuremath{\,\mbox{km}\,\mbox{s}^{-1}}}

\newcommand{\Msun}{$M_{\odot}$}
\newcommand{\HI}{H{\sc i}\ }
\newcommand{\degree}{\ensuremath{^{\circ}}}
\newcommand{\XPOS}{{\sc xpos}\ }
\newcommand{\YPOS}{{\sc ypos}\ }
\newcommand{\PA}{{\sc pa}\ }
\newcommand{\INCL}{{\sc incl}\ }
\newcommand{\VSYS}{{\sc vsys}\ }

\newcommand{\MLstar}{\ensuremath{\Upsilon_{\star}}}
\newcommand{\mnras}{MNRAS}
\newcommand{\apj}{ApJ}
\newcommand{\aj}{AJ}
\newcommand{\aap}{A\&A}
\newcommand{\apjl}{ApJL}
\newcommand{\araa}{ARA\&A}
\newcommand{\nat}{Nature}
\begin{document}
\title[High-resolution rotation curve of NGC 6822.]{A High-Resolution Rotation Curve of NGC 6822: A Test-case for Cold Dark 
Matter}
\author[Weldrake, de Blok \& Walter]{D.T.F.~Weldrake,$^1$ W.J.G.~de~Blok$^2$\thanks{Present Address: Department of Physics and Astronomy, Cardiff University, Queen's Buildings, PO Box 913, Cardiff CF24 3YB, United Kingdom}, F.~Walter
$^3$\thanks{Present Address: NRAO AOC,  P.O. Box O,  1003 Lopezville Road, Socorro, NM 87801-0387, USA}\\
$^1$ Research School of Astronomy \& Astrophysics, Mount Stromlo Observatory, Cotter Road, Weston Creek, ACT 2611, Australia\\
$^2$ Australia Telescope National Facility,
PO Box 76, Epping NSW 1710, Australia\\
$^3$ California Institute of Technology,
Astronomy Department 105-24, Pasadena, CA 91125, USA}

\maketitle
 
\begin{abstract}
We present high resolution rotation curves of the local group dwarf
irregular galaxy NGC 6822 obtained with the Australia Telescope Compact
Array. Our best curves have an angular resolution of $8''$ or 20 pc
and contain some 250 independent points. The stellar and gas
components of NGC 6822 cannot explain the shape of the curve, except
for the very inner regions, and NGC 6822 is consequently very dark matter
dominated. There is no evidence for the presence of a steep density
cusp down to scales of $\sim 20$ pc, contrary to the predictions of
Cold Dark Matter.
\end{abstract}


\section{Introduction}

NGC~6822 is a dwarf irregular local group member. Located at a
distance of $490\pm 40$ kpc \citep{mateo98}, NGC 6822 is the most
nearby dwarf irregular apart from the LMC/SMC system. Due to the small
distance the galaxy appears very extended on the sky: its optical
angular diameter is over a quarter of a degree; the H{\sc i} disk
measures close to a degree \citep{dBW2000}.  NGC~6822 is a member of
an extended cloud of irregulars \citep{mateo98} known as the ``Local
Group Cloud''. The galaxy has a total luminosity of $M_B\!=\!-15.8$
\citep{hodge91} and a total H{\sc i} mass of $1.3
\times 10^8\ M_{\odot}$ \citep{dBW2000}, making it relatively 
gas-rich. It is a metal poor galaxy, with an ISM abundance of about
0.2 $Z_{\odot}$ \citep{evan89} and has a star formation rate of $\sim
0.06\ M_{\odot}{\rm yr}^{-1}$ (based on H$\alpha$ and FIR fluxes)
\citep{mateo98,israel}. \citet{hodge80} found evidence for increased star 
formation between 75 and 100 Myr ago, while \citet{gallart2} showed
that the star formation in NGC 6822 increased by a factor of 2 to 6
between 100 and 200 Myr ago. This is consistent with the mostly
constant but stochastic recent star formation histories often derived
for dwarf and LSB galaxies \citep{grebel,jeroen}.  NGC 6822 can be
regarded as a rather average and quiescent dwarf irregular galaxy.

\citet{dBW2001} found that the outer \HI disk is dominated by what
appears to be a super-giant \HI shell with extended, apparently tidal
features in the outer disk.  In contrast, the inner main \HI disk is
remarkably uniform and circularly symmetric. There is no evidence for
large kinematical disturbances, as we will show in this paper. 

In this paper we address the detailed dynamics and kinematics of NGC 6822
and derive high-resolution rotation curves of this dwarf galaxy.  The
physical resolution of our highest resolution curve is $\sim~20$ pc or $\sim 0.03h$ where $h$ is the exponential scale length of the stellar disk (see Sect.~\ref{sec:opt}),
making it the highest resolution \HI rotation curve of any undisturbed
dwarf galaxy.  

In Section~2 we summarize the ``small-scale crisis'' in current
cosmological simulations.  Section~3 briefly describes the
observations and data reduction, while Sect.\ 4 and 5 describe the
derivation and analysis of the rotation curves in detail.  In Section 6
we describe the models used in rotation curve fitting, while Section 7
presents the results. Section 8 discusses the results, and tries to
fit them in the CDM picture. We summarize our conclusions in Section
9.

\section{Dark Matter in Dwarfs}

Cosmological numerical Cold Dark Matter (CDM) simulations predict a
specific and universal shape for the dark matter mass density
distributions (and therefore the rotation curves) of galaxies
\citep{dubinski}.  This was investigated in detail by
\citet{NFW96,NFW97} who found that the mass-density
distribution in the inner parts of simulated CDM halos could best be
described by a $r^{-1}$ power-law. This steeply increasing density
cusp towards the center naturally translates into a steeply rising
rotation curve.  This kind of curve is however not observed in
late-type disk galaxies \citep[e.g.][]{swaters_phd,marc_phd97} . The
rotation curves of dwarf and Low Surface Brightness (LSB) galaxies are
shallow and rise linearly, more consistent with a dark matter
distribution that is dominated by a constant-density core with a size
of a few kpc
\citep{moore94,edb_rot,paper3,blais00}.  
Especially in the case of LSB galaxies there has been much discussion
about the reality of the observed shallow curves \citep{SMT}.  It was
argued that systematic resolution effects (``beam-smearing'') could
hide the steep CDM curves and lead to an erroneous conclusion that
they would not be consistent with CDM.

Subsequent high-resolution follow-up studies, measuring the rotation
curves of LSB galaxies in H$\alpha$ at a resolution of $\sim 0.2$ to
$\sim 1$ kpc have now confirmed the conclusions derived from the early
data 
\citep{paper3,optcur_data,optcur_model,bosma02,blais00,blitz}
Dwarfs and LSB galaxies seem to be dominated by a dark matter
distribution that is best described in the inner parts by a powerlaw
$\rho \sim r^{\alpha}$ with $\alpha = -0.2 \pm 0.2$ \citep{paper3}.

With a spatial resolution of only $\sim 20$ pc, the current data set
enables us to measure the rotation curve of this dwarf galaxy at a
resolution which is an order of magnitude better than the observations
described above. Among other things, the high resolution makes it
possible to investigate the effect beam size has on mass models and
test whether the inferred distribution of dark matter depends on
resolution.

\section{Observations and data reduction}

NGC\,6822 was observed with the Australia Telescope Compact Array for
$15 \times 12$ hours in its 375 (1 $\times$ 12h), 750D ($2 \times
12$h), 1.5A (4 $\times$ 12h), 6A and 6D ($8 \times 12$h)
configurations over the period from June 1999 to March 2000. A total
of 8 pointings was observed covering the entire H{\sc i} extent of the
galaxy.  We used a bandwidth of 4 MHz with a channel separation of 0.8
\kms.  Additionally, to obtain zero-spacing information, NGC 6822 was 
observed with the Parkes single dish radio telescope using the
multibeam receiver in its narrowband mode in December 1998. The
correlator configuration was identical to that used for the Compact
Array observations.

The data were reduced and mosaicked together using the {\sc miriad}
data reduction package.  Super-uniform weighting, reducing side lobes
in individual pointings prior to mosaicing, was used. The resulting
data cubes were cleaned with the {\sc miriad mossdi} task.  We
combined the single dish and synthesis data to correct for the missing
zero-spacings and produced data cubes at various spatial and velocity
resolutions. For the rotation curve analysis presented here we use the
data cubes with channel separations of 1.6
\kms\ (with an effective velocity resolution of 1.9 \kms). These cubes
gave the best compromise between velocity resolution and
signal-to-noise.

We analyzed the 1.6 \kms\ data at five different spatial
resolutions. The synthesized beam sizes used were $96''\times
349.4''$, $48'' \times 174.7''$, $24'' \times 86.4''$, $12'' \times
42.4''$ and $8'' \times 28.3''$.  The position angle of the beam was
0$^\circ$ at all resolutions. We will refer to the 5 different
resolution data sets as B96, B48, B24, B12 and B08,
respectively. Table~1 gives some more information on the data sets. We
retained the elliptical beamshape to obtain the highest possible
resolution along the major axis. As the PA of the major axis of NGC
6822 is $\sim 110\degree$, as shown later, this means that the
resolution along the major axis is to a few percent equal to the minor
axis beam size.  The large size of the galaxy compared to the beam
rules out any beam smearing effects due to non-major axis information
entering the beam.

\section{Moment maps and velocity field}

All subsequent analysis was done using the {\sc gipsy} package. The
low-resolution cube B96 was clipped at the $2\sigma$ level, and
remaining noise peaks were removed by hand.  Then for B48 and
subsequent resolutions, we used the $2\sigma$-clipped cube of the
previous resolution as a mask, after which remaining spurious noise
peaks were removed by hand. This ensures a consistent selection of
features at all resolutions. Furthermore using the lower resolution
cube as a mask ensures that possible extended low-level structures are
retained in the map.

\begin{figure*}
\psfig{figure=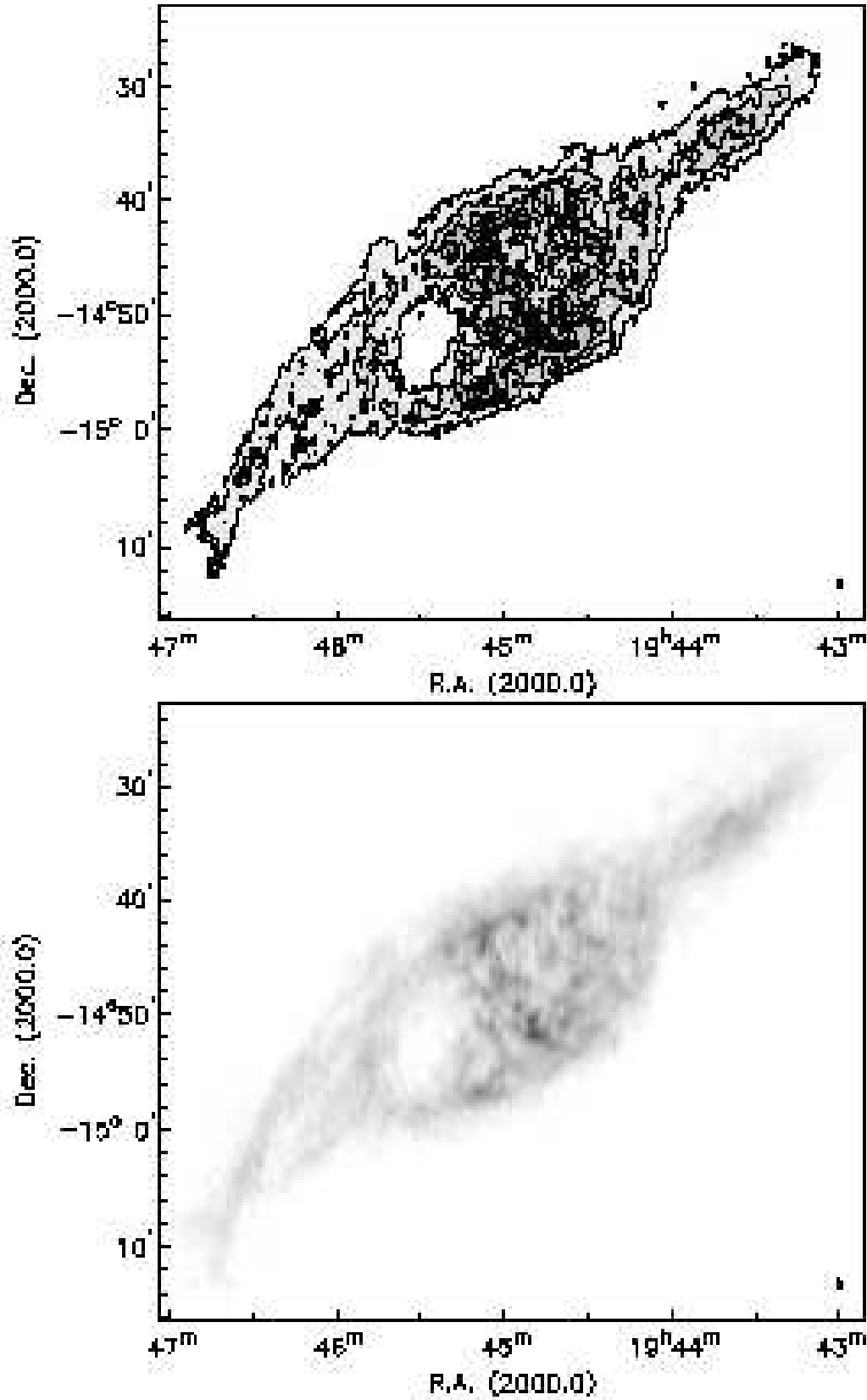,width=0.8\hsize}
\caption{B12 total column density map. Contours start at 5 $M_{\odot}$ pc$^{-2}$ and increase in steps of 1 $M_{\odot}$ pc$^{-2}$. The beam is indicated in the lower right corner.
\label{b12mom0}}
\end{figure*}

The integrated \HI surface density maps were made in the usual manner by
adding together all clipped channel maps.  As a second step we then
isolated the high signal-to-noise (S/N) regions of the maps as follows. 
For uniformly tapered maps in velocity $\sigma_{tot}=
\sqrt{N}\sigma_{ch}$, where $\sigma_{tot}$ is the noise in a pixel in
the integrated column density map, $N$ is the number of channels
contributing to that pixel, and $\sigma_{ch}$ is the noise in one
channel at that pixel.  We constructed noise maps for each integrated
column density map, and used these together with the column density maps
to isolate those pixels in the column density maps where $S/N > 10$. 
These high $S/N$ maps were used as masks for the velocity fields. 
Figure~\ref{b12mom0} shows the integrated column density map derived for
the B12 data. 

There are two standard ways to produce a velocity field.  One commonly
used procedure is to determine the intensity weighted mean of the
velocities along each profile. It is known that this method can
produce spurious velocities at low resolution and low
S/N ratios. Given the high resolution and S/N of
our data set, this is unlikely to be a problem here.  An alternative
method is to fit (a) Gaussian(s) to each profile. The latter method
usually gives better results, as it is less affected by systematic
effects due to skewed profiles etc. It is however computationally more
expensive which becomes important for data cubes of the size we are
dealing with here. The high quality of the current data puts us firmly
in the regime where the intensity weighted mean gives accurate
results.

\begin{figure*}
\psfig{figure=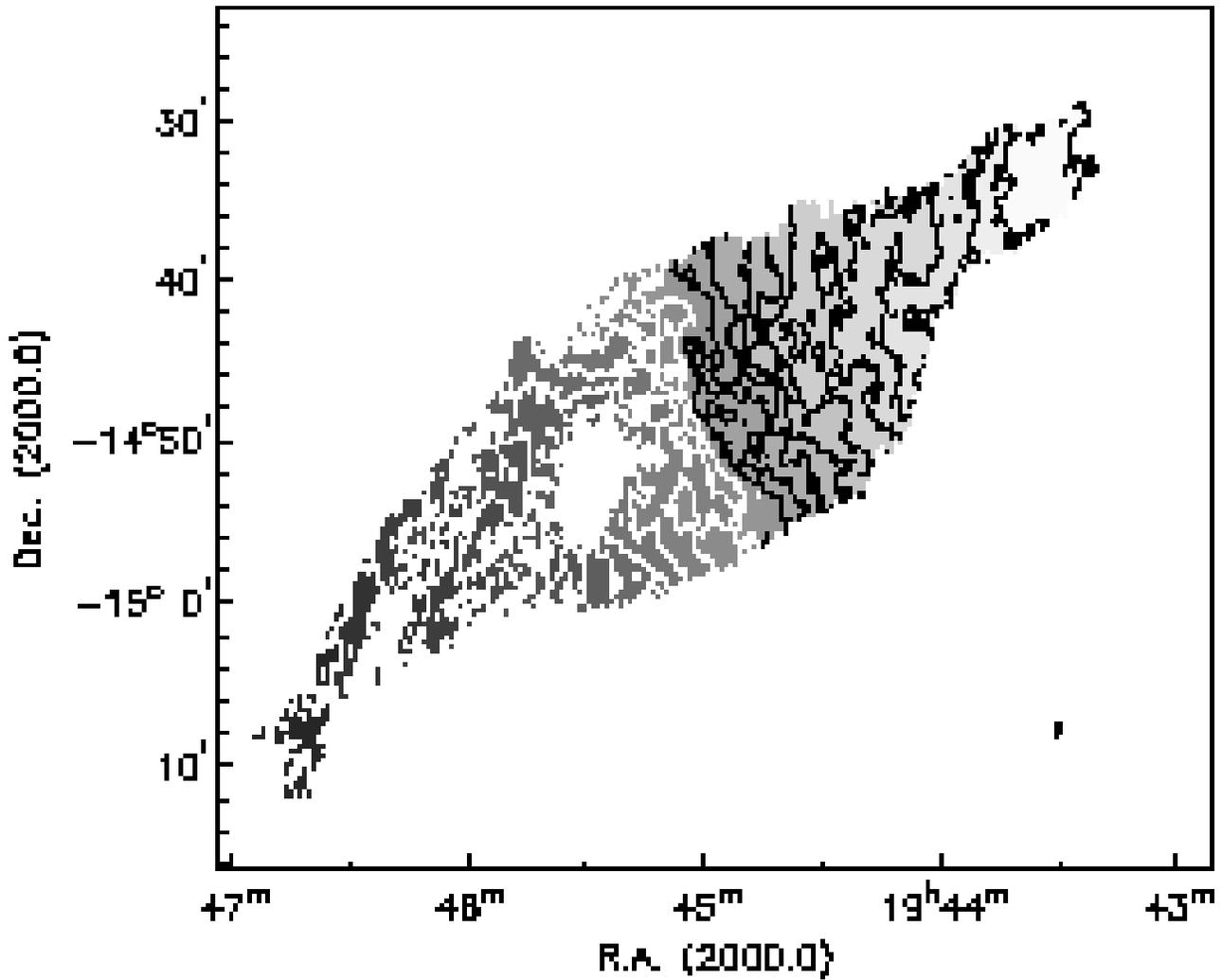,width=\hsize}
\caption{The intensity weighted velocity field, derived from the B12 data. 
The black contours run from $-55$ \kms\ in the central parts to $-100$
\kms\ in the outer NW parts, in steps of 5 \kms. The white contours
run from $-50$ \kms\ in the inner parts to $+10$ \kms\ in the outer SE
parts.  The beam is shown in the lower-right
corner.\label{compvelfie}}
\end{figure*}

To test this we have compared both methods for the B12 data set
(Fig.~\ref{compvelfie}) and find that for our data the differences are
negligible. A histogram of the differences is very well described by a
Gaussian with an average of $-0.3$ \kms\ and an RMS of $1.0$
\kms, i.e., both methods produce identical results to 
within better than a channel spacing.  The number of pixels where the
residuals reach $5$ \kms\ or higher is only $\sim 0.1$ percent of the
total number of pixels.

\begin{figure*}
\psfig{figure=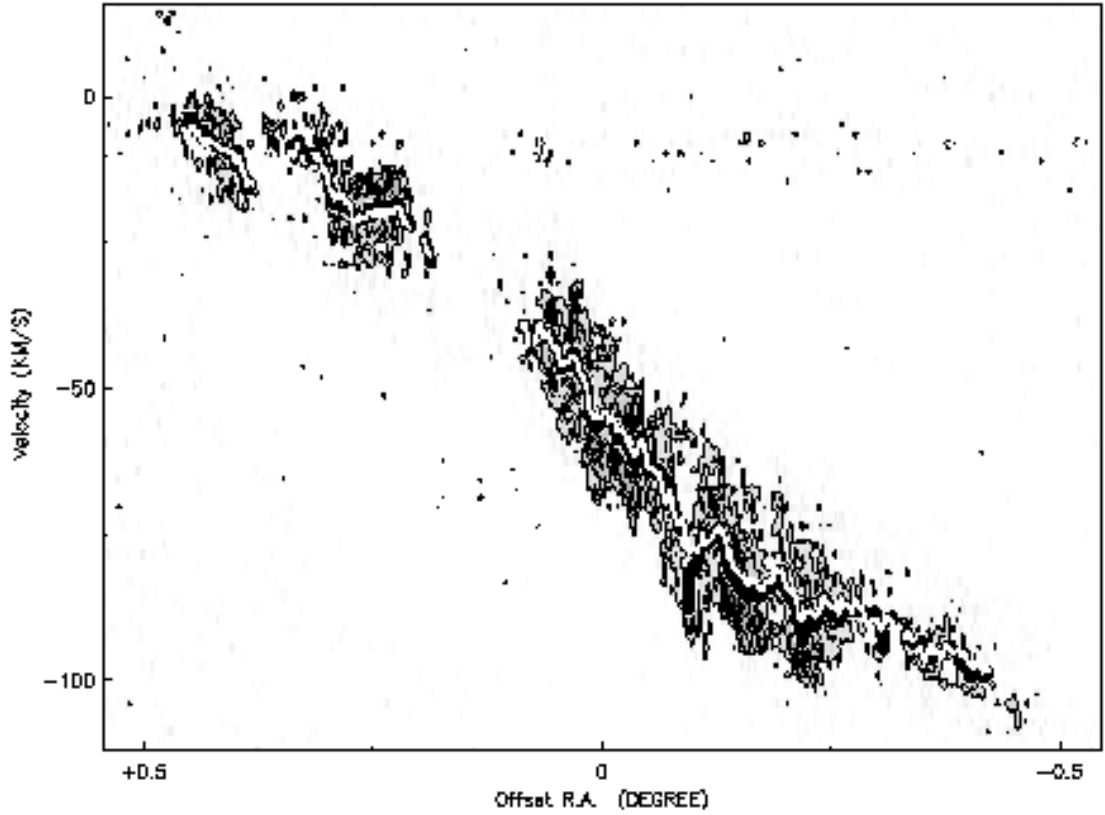,angle=-90,width=\hsize}
\caption{B12 data major axis position-velocity diagram taken along position angle 
120\degree. Dotted contour represents $-4\sigma$, full contours start
at 4$\sigma$ and increase in steps of 2$\sigma$. The white line shows
the values found in the intensity weighted velocity field. The black
line shows values derived from the gaussian velocity
field.\label{majaxslice}}
\end{figure*}

Figure~\ref{majaxslice} shows a major axis position velocity diagram
of the B12 data, taken along a position angle of 120\degree.
Over-plotted are the velocities as found in the intensity weighted and
Gaussian velocity fields. The velocity fields produce the same
results, without large-scale systematic deviations.  We will therefore
proceed to use intensity weighted velocity fields in subsequent
analysis

\section{Deriving Rotation Curves}

The rotation curves were produced using the {\sc gipsy} task {\sc
rotcur}.  For a well-resolved and high S/N data set such as the
current one, the tilted ring procedure is by far the best way to
determine the rotation curve. Other methods which are often used, such
as adjusting the tilted ring parameters by hand using position
velocity diagrams as a guide (e.g.\ using the {\sc gipsy} command {\sc
inspector}), are more subjective and results can depend on
guesses on e.g.\ the magnitude and effects of beam
smearing. 

A look at Fig.~\ref{compvelfie} shows that the high-resolution
velocity fields contain an extra-ordinary amount of detail. 
To get a good feel for the large scale structure in NGC 6822 we started
by deriving the B96 rotation curve, and used the results for each
resolution as the initial estimate for the next higher resolution,
thus gradually refining the curve.

We adopted rings with a width of equal to the minor axis of the
beam. The beam is elongated roughly along the minor axis of NGC 6822 with
an effective radius of $\sim 1.9$ times the minor axis. A ring
width equal to the minor axis is a good compromise between beam shape
and complete sampling.  Most information contributing to the rotation
curve comes from near the major axis where the beam is narrow.

In general the procedure involved making a fit with all parameters
(systemic velocity, centre position, position angle [PA], inclination and
rotation velocity) free.  The central position and systemic velocity
were then fixed and several runs with either
inclination or PA or both fixed were made to find the best model.

As the minor axis generally provides little information regarding the
rotation curve we excluded an angle of 30\degree\ around the minor
axis from the fits. One usually also down-weights the data around the
minor axis by applying a $|\cos \theta |$ weighting, where $\theta$ is
the angle with respect to the major axis in the plane of the galaxy.
Other alternatives are to apply a uniform weighting or a $\cos^2
\theta$ weighting.  We have experimented with different weighting
schemes as well as varying the exclusion angle around the minor axis
between 15\degree\ and 45\degree, but found no difference in the curves
produced (this is due to the large number of independent points that
{\it do} contribute at full weight around the major axis). We
therefore adopt a uniform weighting scheme and a free angle around the
minor axis of 30\degree.

Once a satisfactory curve using both sides of the velocity field was
produced, we also calculated two curves using only the approaching and
receding sides of the velocity field. Comparing the three curves gives
information about the symmetry of the system.

\subsection{The rotation curves}

\begin{figure*}
\begin{center}
\begin{tabular}{ccc}
\psfig{figure=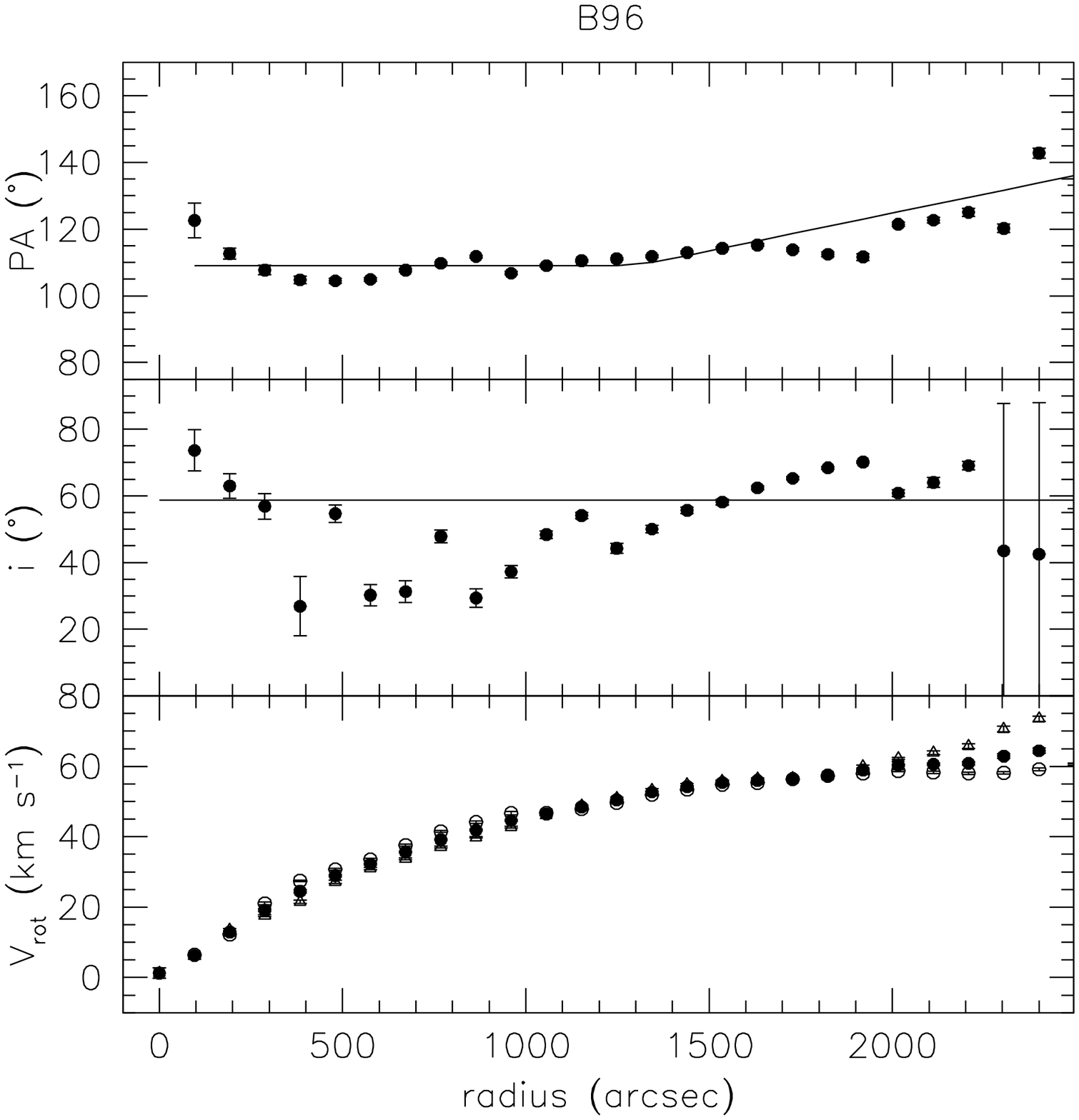,width=0.35\hsize}&
\psfig{figure=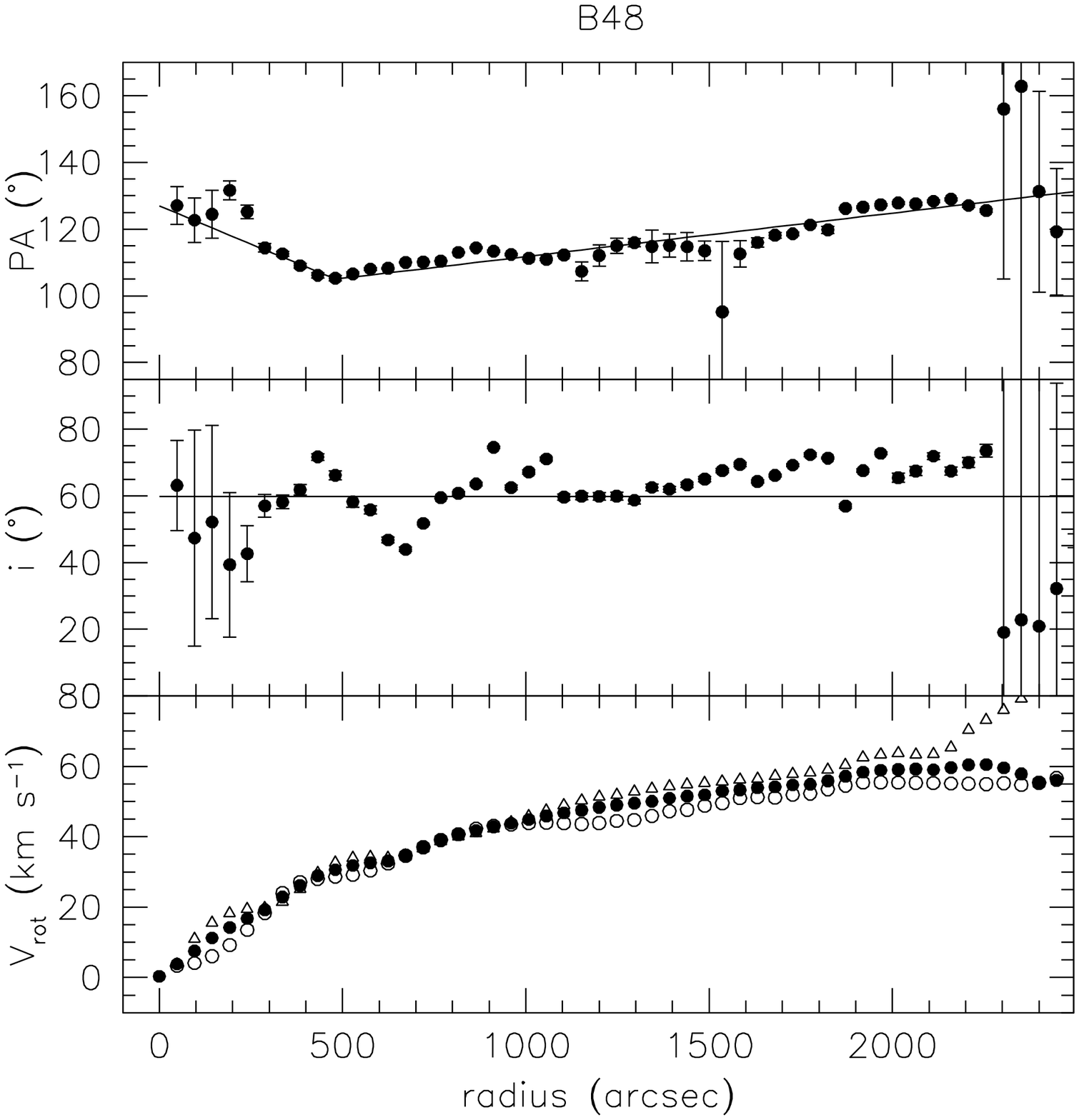,width=0.35\hsize}\\
\psfig{figure=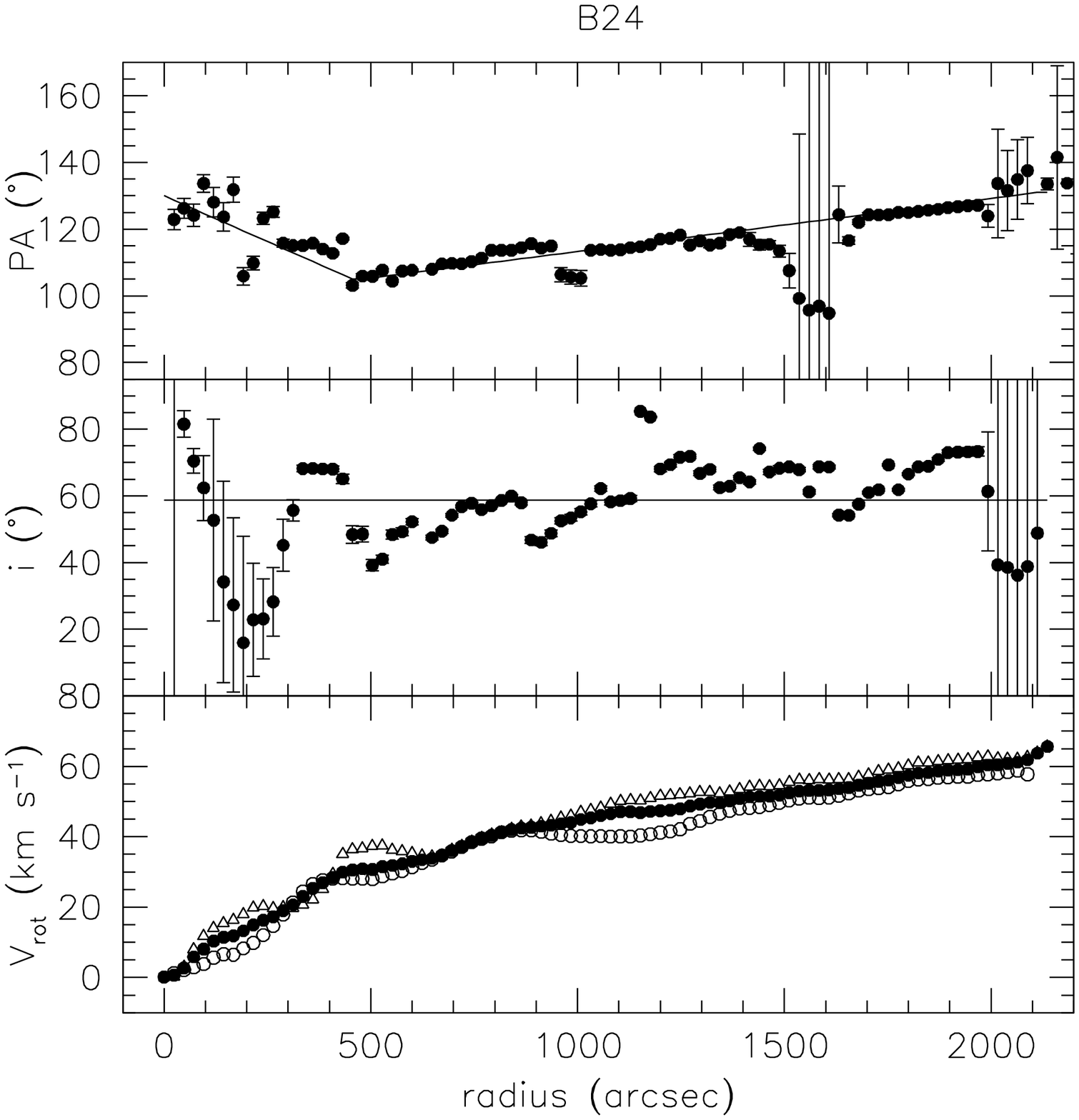,width=0.35\hsize}&
\psfig{figure=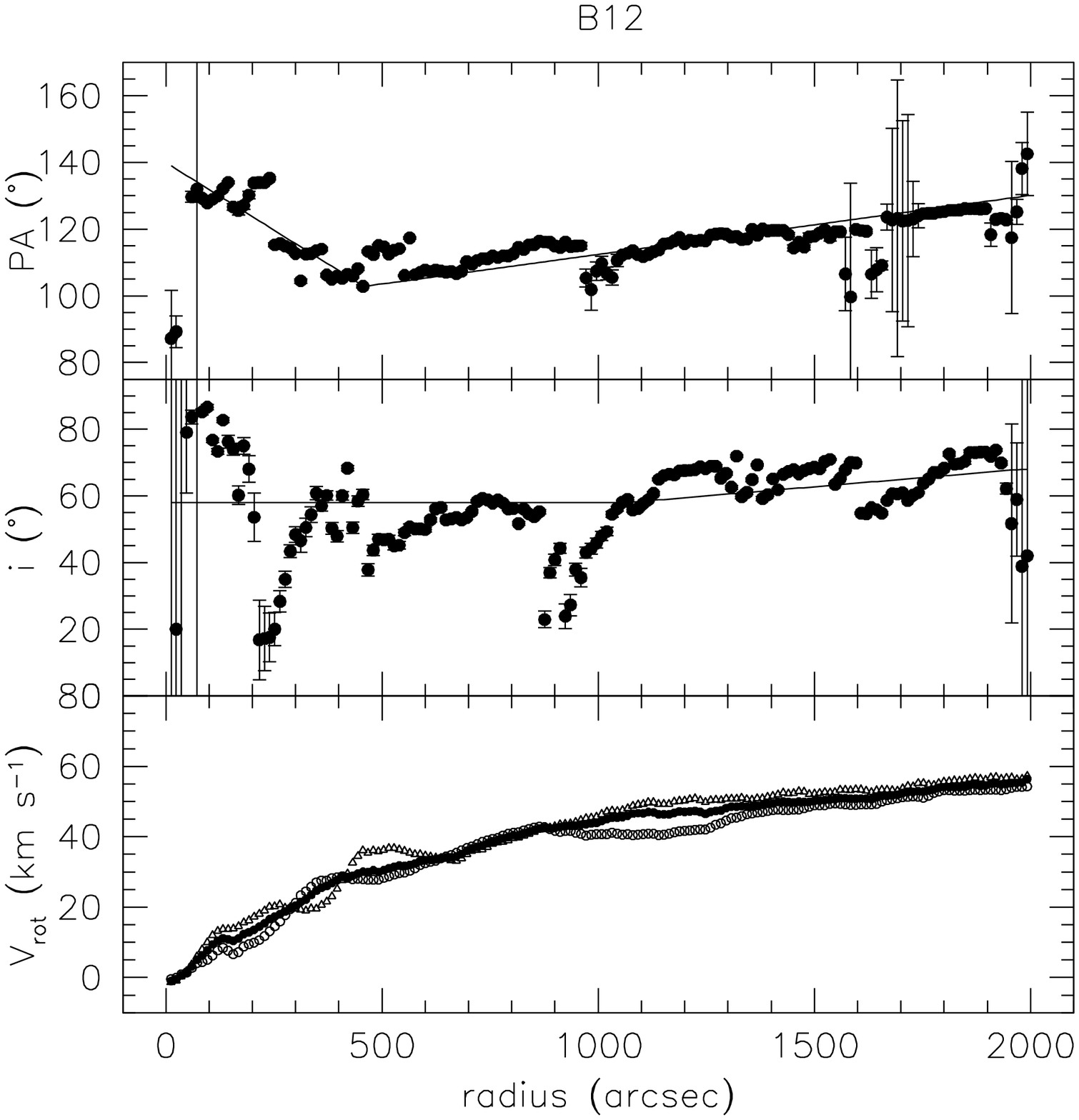,width=0.35\hsize}\\
\psfig{figure=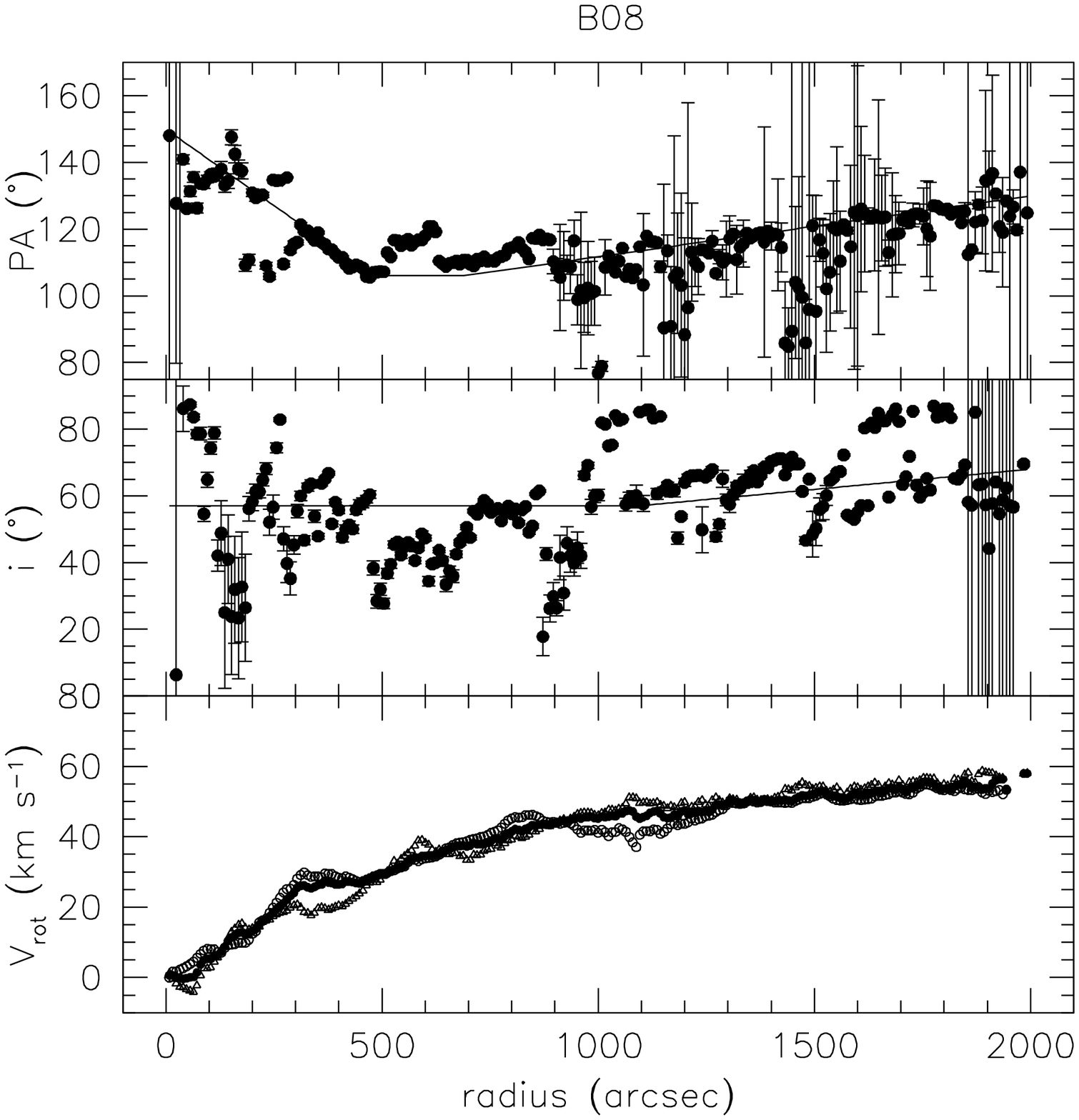,width=0.35\hsize}&
\psfig{figure=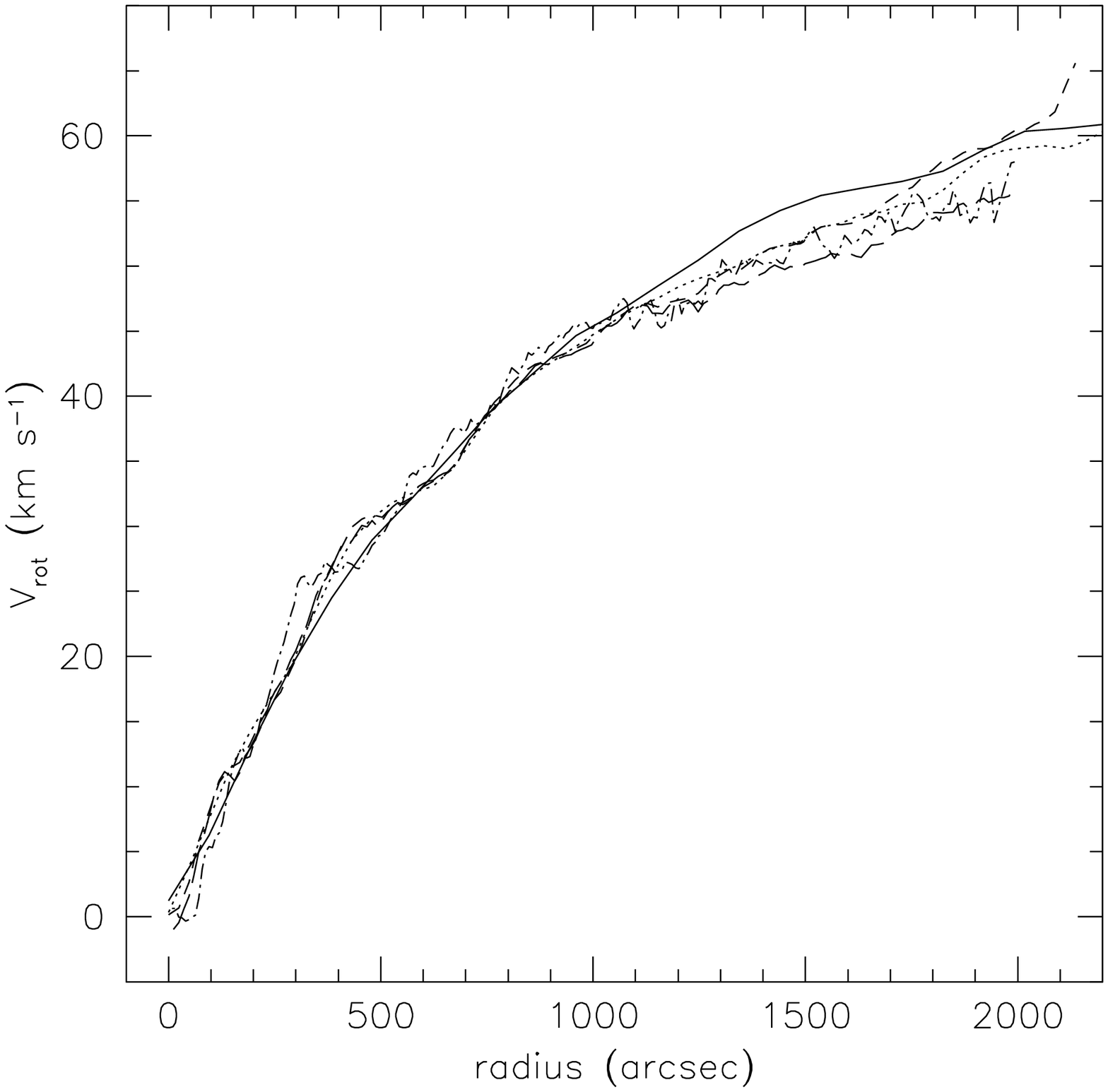,width=0.35\hsize}\\
\end{tabular}
\end{center}
\caption{Results of the tilted ring fits for the B96, B48 (top row), B24, B12 (middle row) and B08 (bottom-left) resolutions.
In each panel we show the 
the position angle of the major axis (top), the inclination 
(centre) and circular velocity (bottom) as a function of radius for
the various resolutions.  In every top sub-panel the
filled circles represent the curve derived using the whole velocity
field. The open circles represent the curve for the approaching (SE)
side, while the triangles represent the receding (NW)
side. Overplotted in the top and centre sub-panels are the models used to
derive these curves. The points in the top sub-panel show the behaviour of
PA as a free parameter, with inclination fixed to its model value as
shown in the center panel. The points in the center sub-panel show the
behaviour of inclination with PA fixed to the model values as shown in
the top panel. The bottom-right panel shows the rotation curves at each resolution overplotted. Full line: B96; dotted line: B48; short-dashed line: B24; long-dashed line: b12; dash-dotted line: B08.\label{b96painclvrot}}
\end{figure*}

For each resolution we derived the curve using the procedure described
above.  The results are shown in Table~2 and
Fig.~\ref{b96painclvrot}. The trends found in PA and inclination for
the different resolutions agree well with each other.  Only the B96
data does not show the PA trend as strongly.  The variation of
inclination with radius is only small; only the highest resolutions
show weak evidence of a slight warp in the outer parts: choosing a
constant inclination instead therefore does only marginally affect the
curves. The trend in PA is pronounced and real; deriving curves with a
constant PA results in velocities and inclinations that are
inconsistent with the data, especially at the higher resolutions.  The
small-scale kinks in the inclination are not physical and  mainly
due to dispersion effects.

Beyond $R\sim 1000''$ (2.5 kpc) the difference between the approaching
and receding sides becomes more pronounced with increasing resolution.
This radius corresponds to the edge of the inner \HI disk. The
receding curve shows no significant features there, but the
approaching curve shows a dip in velocity. This coincides with the
interface between the main \HI disk and the NW cloud \citep{dBW2001}.
However, at larger radii the receding curve converges with the
approaching one, indicating that the galaxy and cloud (NW side) as
well as the ``tails'' (SE side) are embedded in one symmetrical halo.
The global dynamics of the system appear undisturbed despite the
morphology of the \HI. The small kink at $R\sim 150''$ (0.38 kpc) is
caused by a number of high velocity dispersion regions that over a
small radial range happen to be aligned with the tilted rings.  At
$R\sim 500''$ a similar small kink corresponds with the inner edge of
the large hole.

\begin{figure*}
\begin{center}
\begin{tabular}{cc}
\psfig{figure=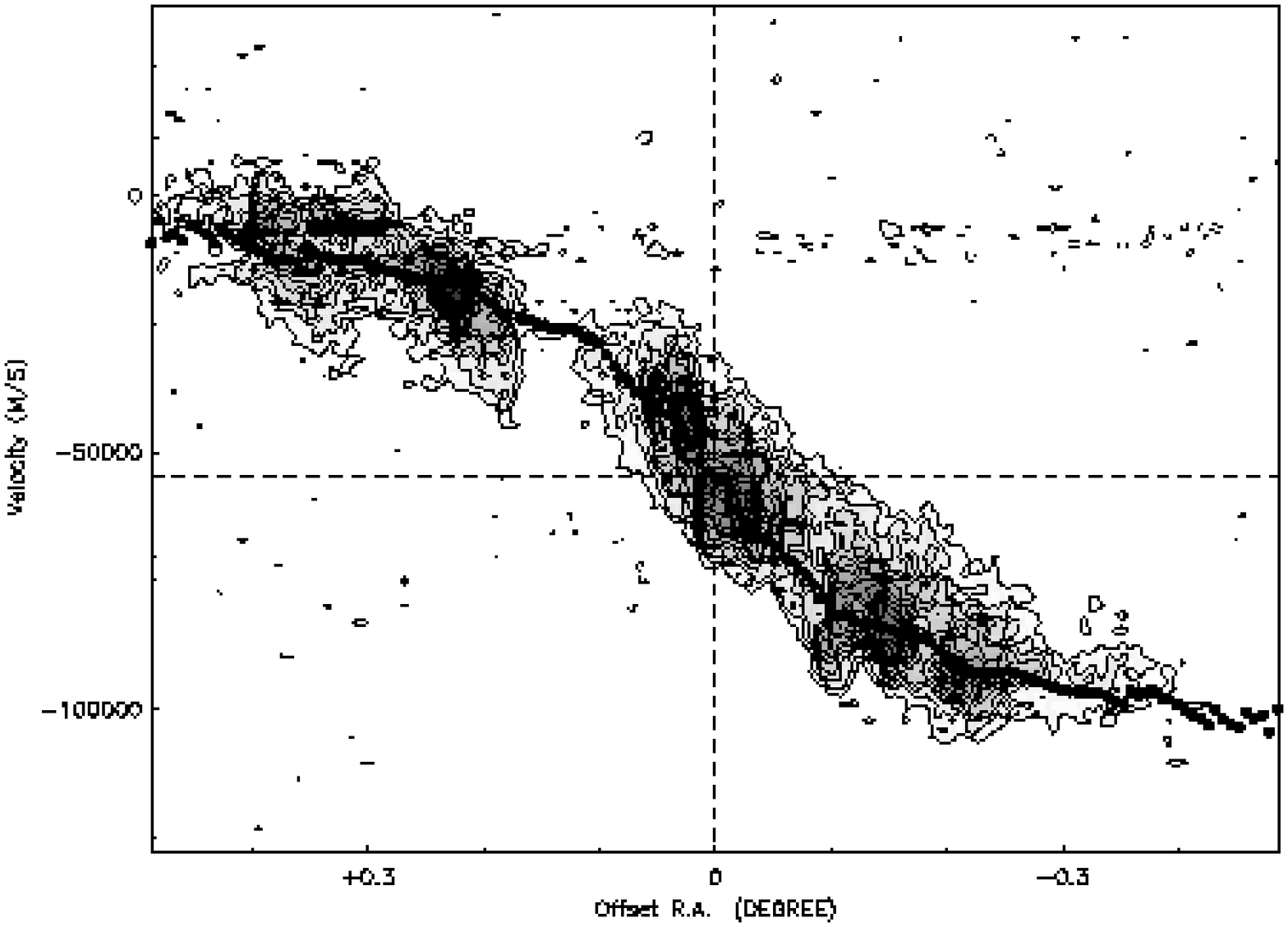,angle=-90,width=0.49\hsize}&
\psfig{figure=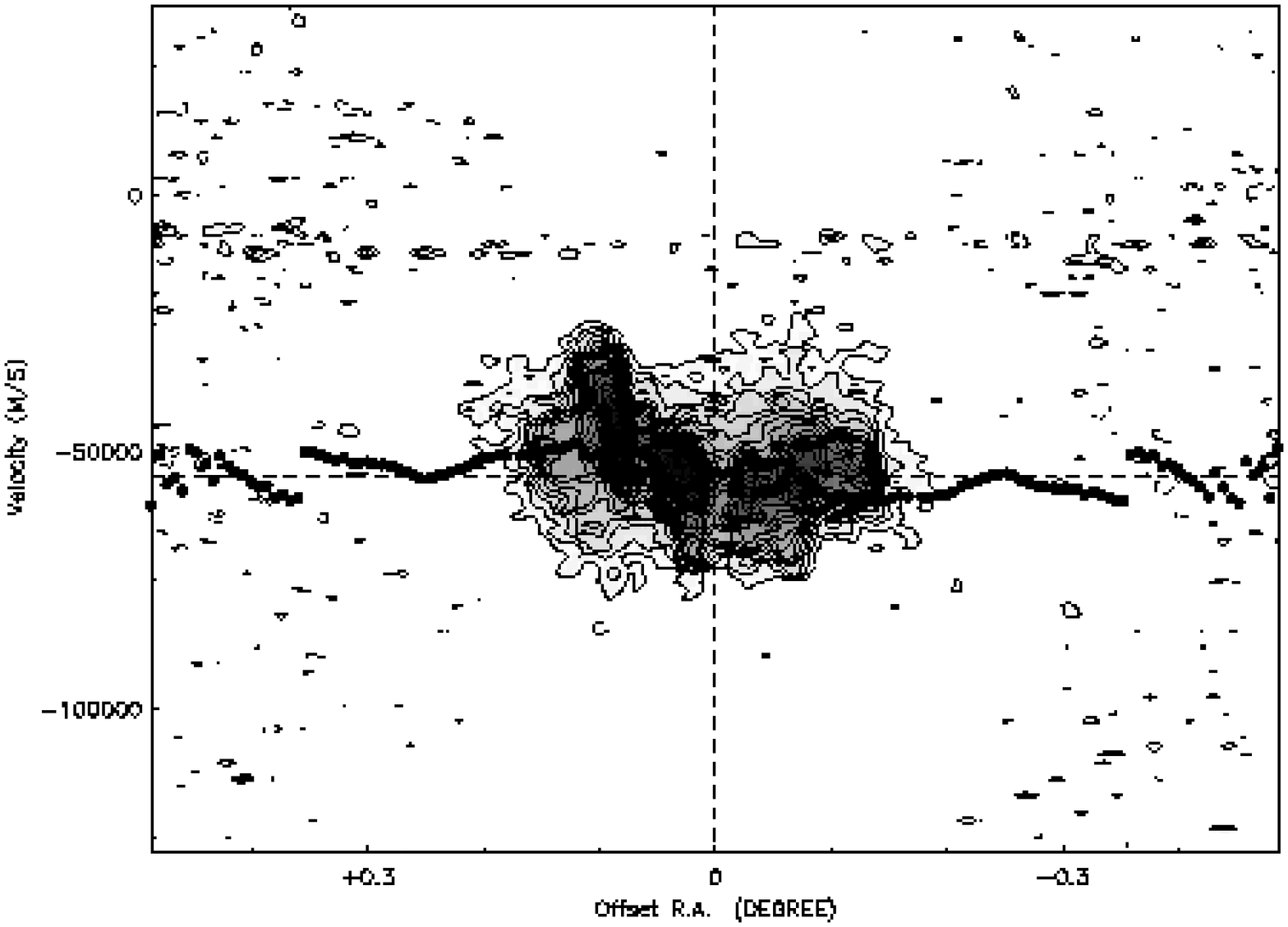,angle=-90,width=0.49\hsize}\\
\psfig{figure=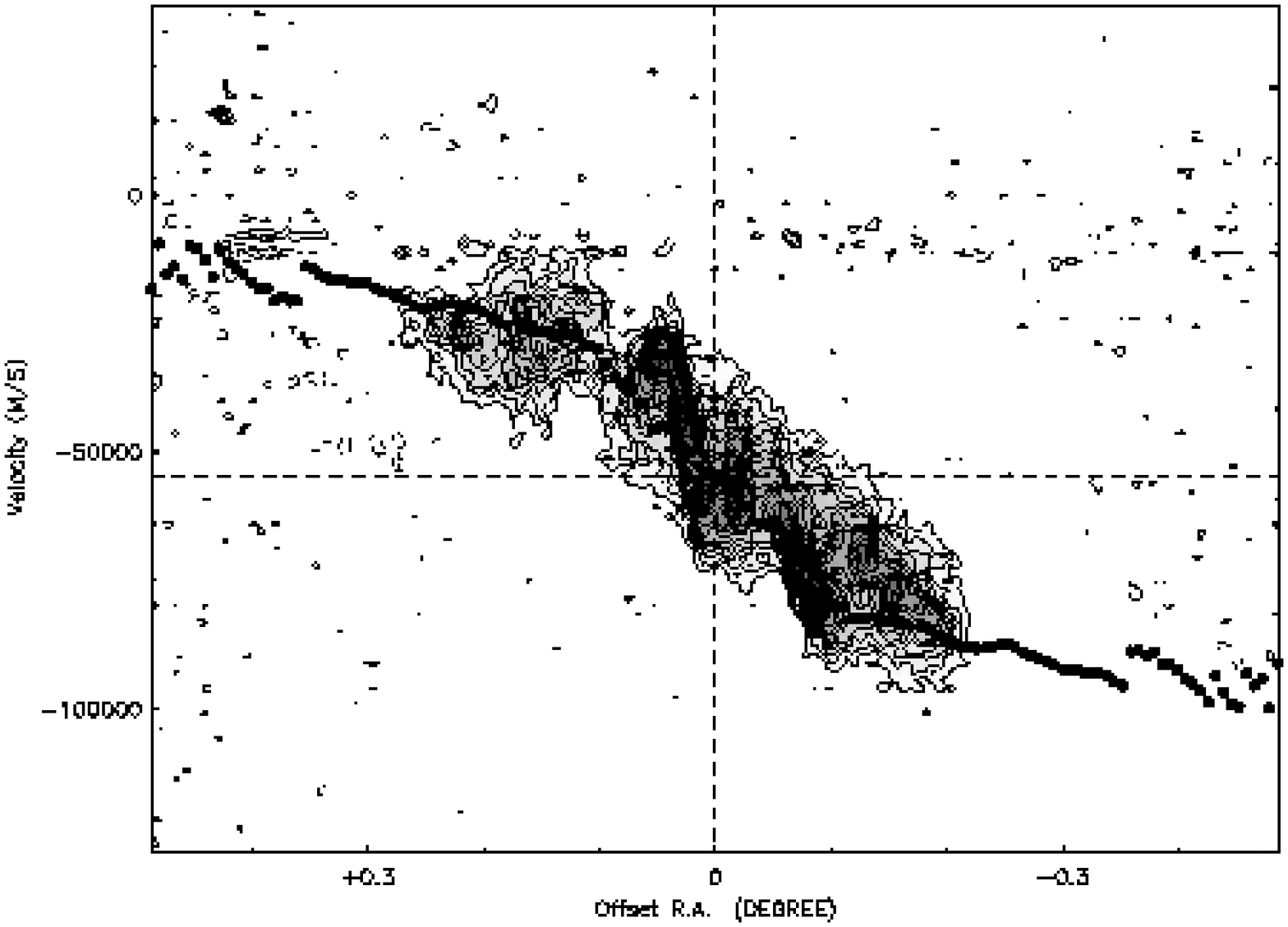,angle=-90,width=0.49\hsize}&
\psfig{figure=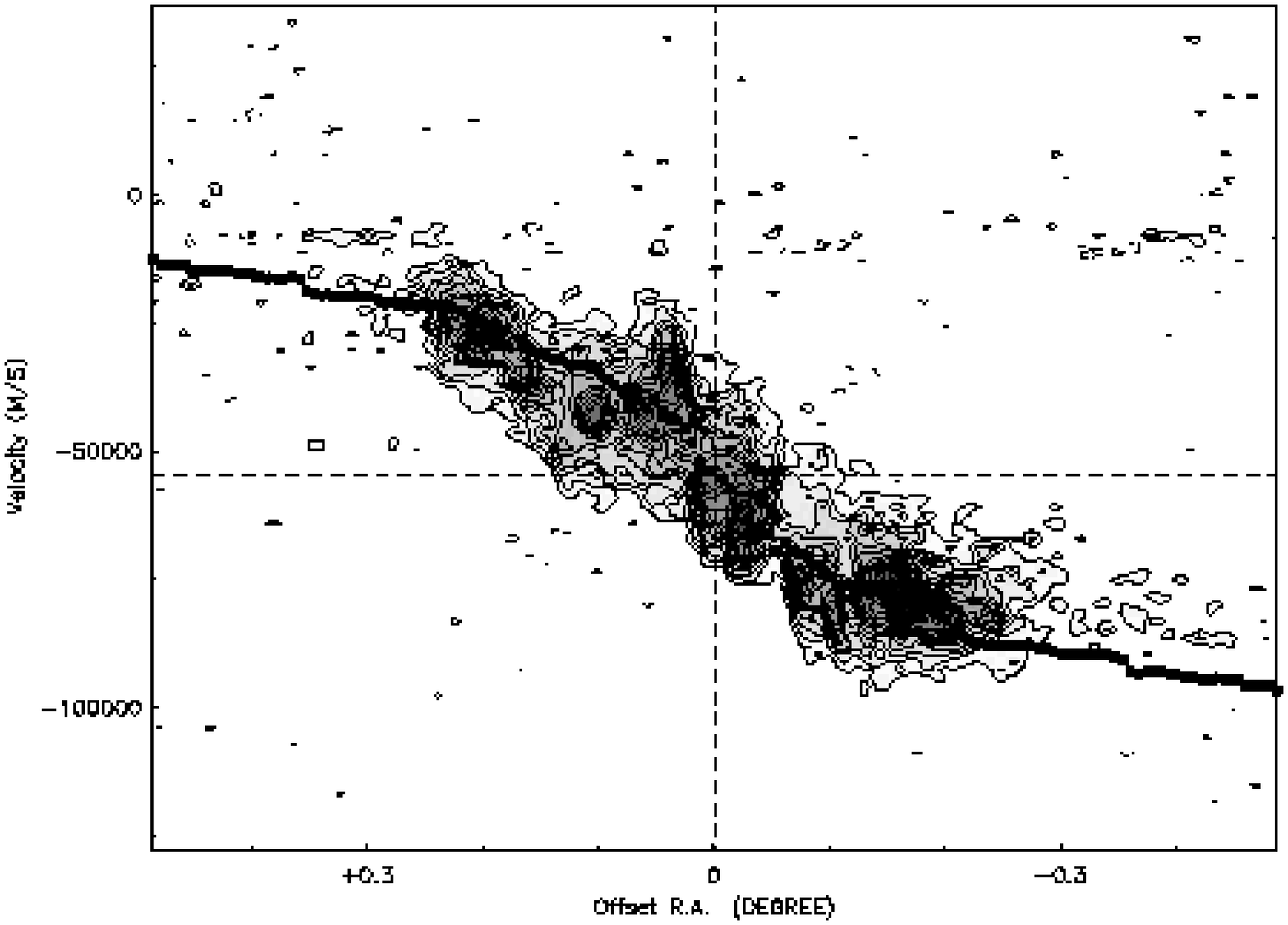,angle=-90,width=0.49\hsize}\\
\end{tabular}
\end{center}
\caption{Position velocity slices of the B24 data, with the 
projected rotation curve overplotted.  The full contours are plotted
starting at 3$\sigma$ in steps of 2$\sigma$. The dotted contours show
the $-3\sigma$ and $-5\sigma$ levels. There is residual Milky Way
emission present at $\sim -10$ \kms. The dotted lines represent the
center of the galaxy and the systemic velocity.  Overplotted is the
final curve as derived from the tilted ring fits, corrected to the
position angle of the slice. Top left: major axis position-velocity
slice taken at a positon angle of 115\degree.  Top right: minor axis
position-velocity slice. Bottom left: position angle of major axis
minus $30$\degree. Bottom right: position angle of major axis plus
$30$\degree.  NGC 6822 is, despite the disturbed morphology, a very
symmetrical galaxy. \label{b24majaxisoverlay}}
\end{figure*}

Figure~\ref{b24majaxisoverlay} shows an overlay of the rotation curve
on various position-velocity slices. The rotation
curve is a good representation of the dynamics of the galaxy as a
whole.

\begin{figure*}
\psfig{figure=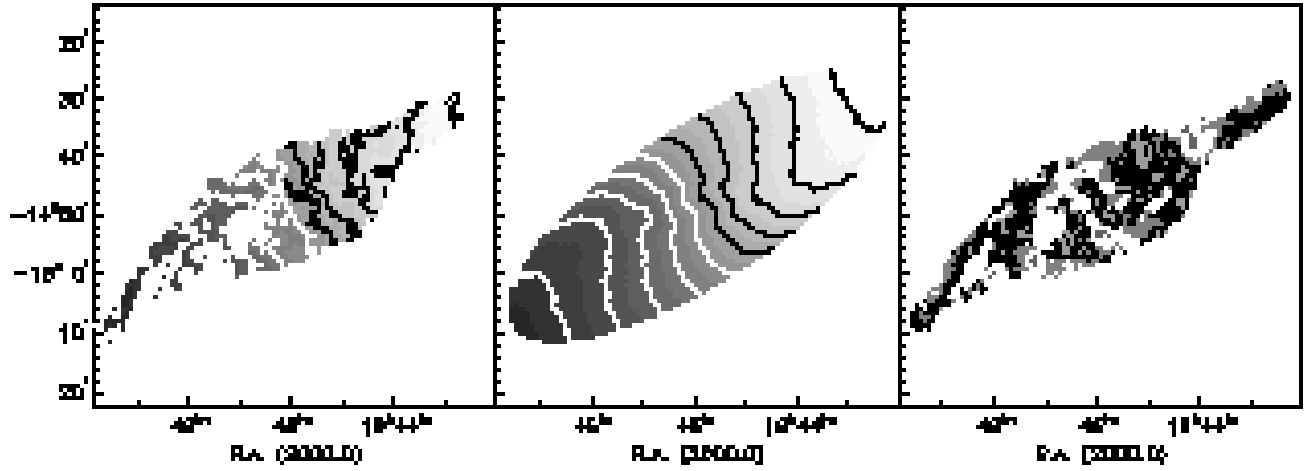,width=\hsize}
\caption{Comparison of the observed velocity fields (left column) and the
circularly symmetric model velocity fields derived from the tilted
ring fits (centre column). The white velocity contours start in the
center of the velocity field at $-50 \kms$ and increase in steps of 10
\kms. The black contours start in the centre at $-55$ \kms and
decrease in steps of 10 \kms towards the NW. The right most column
shows the residual velocity field (observed$-$tilted ring).  The first
white contour represents $+5$ \kms and subsequent white contours
increase with a step of 5 \kms. The first black contour shows the $-5$
\kms and subsequent black contours decrease in steps of 5
\kms.\label{resvelfie}}
\end{figure*}

For all resolutions we have compared the velocity fields with model
velocity fields constructed from the tilted ring fits, and find no
systematic large scale residuals. Figure~\ref{resvelfie} compares the
B12 velocity fields. There are only a few localised regions of
slightly higher residuals in the higher resolution models, coinciding
with regions of high velocity dispersion. Other resolutions give
similar results.

\subsection{Final curves}

The error-bars derived from the tilted ring fits are an indication of
the scatter in velocity around the best-fitting velocity at each
tilted ring.  As such they do not take into account large-scale
asymmetries, differences between approaching and receding sides
etc. To define more realistic errors, we regard the absolute
difference between the rotation velocities of the approaching and
receding sides as $2\sigma$ errors.  We adopt the maximum of the
tilted ring uncertainties and the asymmetry uncertainties as our
error-bars, with one further modification.  In some cases both sides of
the galaxy are very symmetrical, and these usually are also the radii
where the tilted ring uncertainties are very small. These combine to
give very small errors $(\ll 0.1 \kms)$.  We have therefore imposed a
minimum error of half a channel width on the curves. After corrections
for inclination this translates in a minimum error of $0.8$ \kms.

\begin{figure*}
\psfig{figure=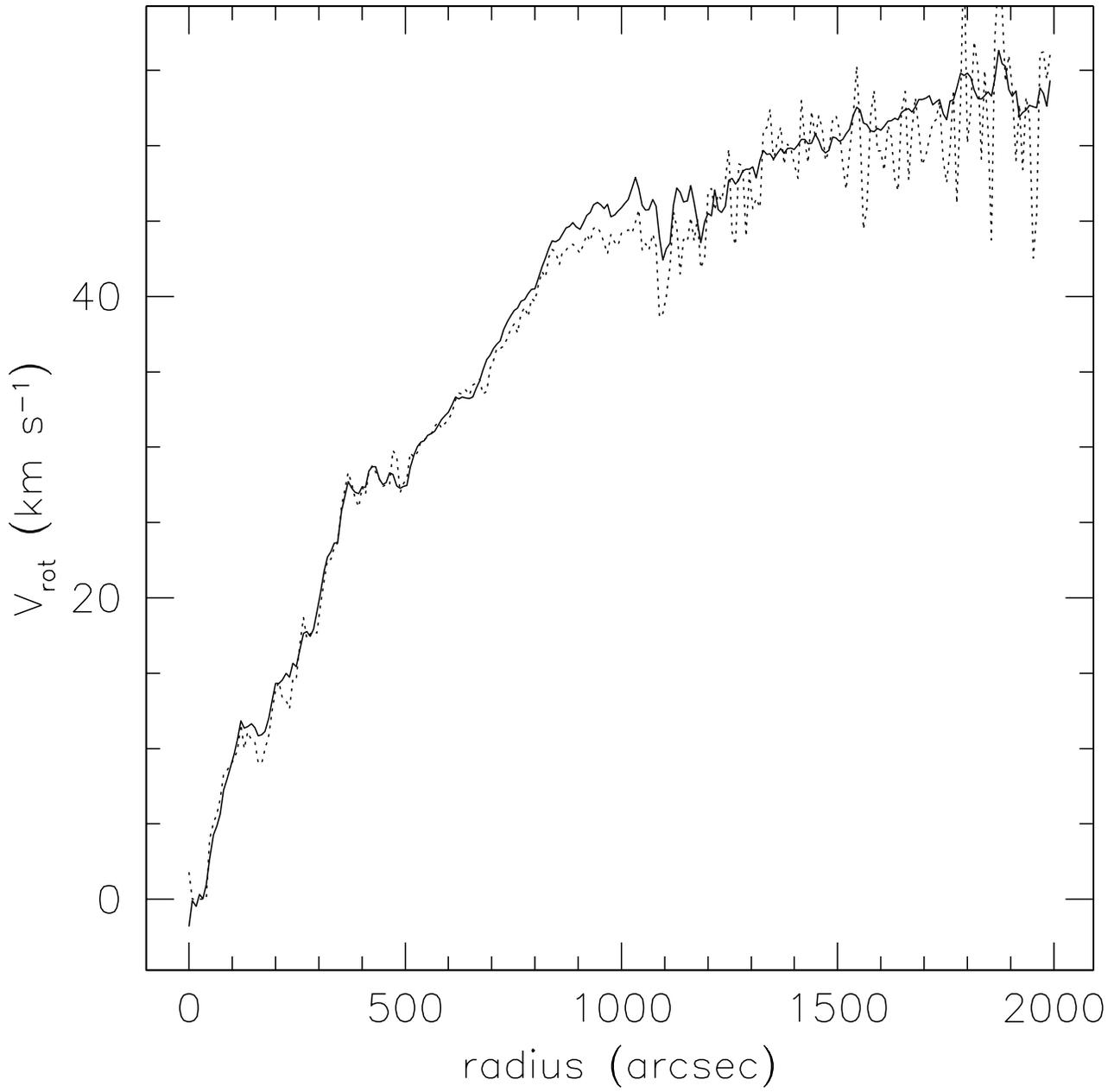,width=\hsize}
\caption{The B08 rotation curve (solid line) and corrected for asymmetric drift (dotted line) \label{assym}}
\end{figure*}

In principle the \HI rotation curves also need to be corrected for the
pressure gradients in the gas to derive the true rotation
velocities. This correction for asymmetric drift is given by $$v_c^2 =
v_{\phi}^2 - \sigma^{2} \left( \frac{\partial \ln \rho}{\partial \ln
R} + \frac{\partial \ln \sigma^{2}}{\partial \ln R} \right),$$ where
$v_c$ is the true rotation velocity, $v_{\phi}$ the observed gas
rotation velocity, $\sigma$ the velocity dispersion in the gas, and
$\rho$ the volume density. The median (modal) velocity dispersion in
the disk of NGC 6822 is $\sim 5.8$ ($\sim 5.7$) \kms, as measured in
the 1.6 \kms\ channel spacing data, and does not show any obvious
radial trends.  Assuming a constant scale height, we can derive the
correction for asymmetric drift. We consider the B08 curves where, due
to the large gradients in the \HI surface density, the corrections are
largest. The corrected and uncorrected curves are shown in
Fig.~\ref{assym}.  Both curves are very similar, with the largest
corrections of about $\sim 3 \kms$ occuring between 850$''$ and
1050$''$.  It should be kept in mind that the correction itself is
also uncertain and depends on assumptions of e.g.\ constant scale
height, which may not hold in the outer parts of the gas disk. As the
uncertainties due to asymmetric drift in the inner parts are
negligible, ignoring them will not affect the results. Based on this
analysis, we decided not to apply this correction.

\subsection{Dynamical centres}

\begin{figure*}
\psfig{figure=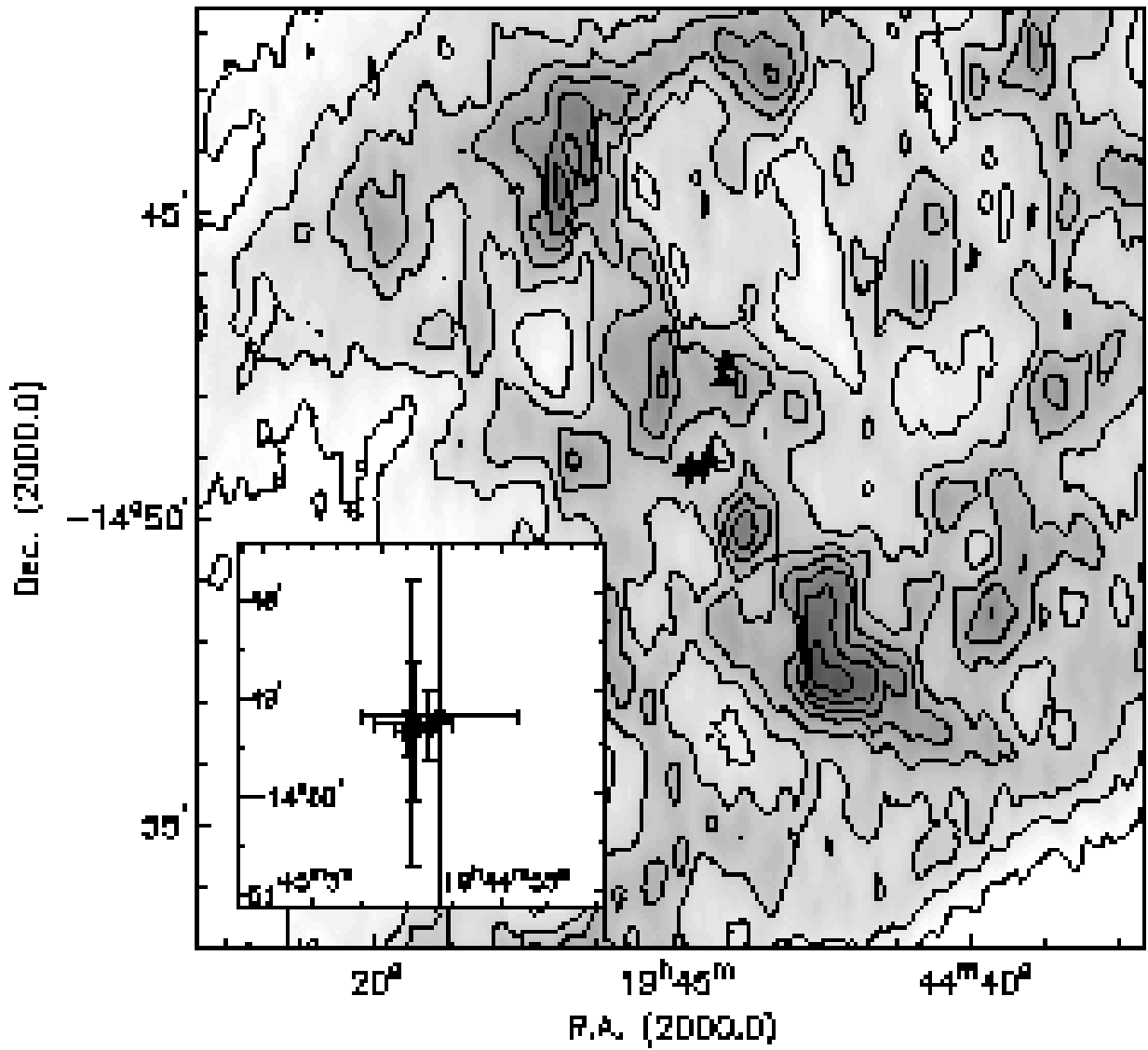,width=\hsize}
\caption{Positions of the dynamical centers of the 5 resolutions (crosses) overplotted on the B12 total column density map. Contours start at 5 $M_{\odot}$ pc$^{-2}$ and increase in steps of 1 $M_{\odot}$ pc$^{-2}$.
The triangles show the positions of the $K$ band isophotes (top) and $R$ band isophotes (bottom). The inset more clearly shows the relative positions of the dynamical centers. Errorbars span the FWHM beam. The centers all coincide to
better than a beam width.
\label{centerpos}}
\end{figure*}

The central position as derived above varies slightly
from resolution to resolution. This is understandable given the
different beam sizes. Fig.~\ref{centerpos} shows the central positions
over-plotted on a B12 column density map of NGC 6822, where the error-bars
span the major and minor axis FWHM of the respective beams.  Also
plotted are $K$-band 2MASS (see below) and $R$-band isophotal
centers. Due to the large size, low galactic latitude and the
irregular morphology the optical center in NGC 6822 is difficult to
determine.  From different fits performed at various isophotal levels
we estimate an uncertainty of $\sim 0.5'$.  These positions are offset
from the dynamical center by about 105$''$ or $0.25$ kpc.

\begin{figure*}
\psfig{figure=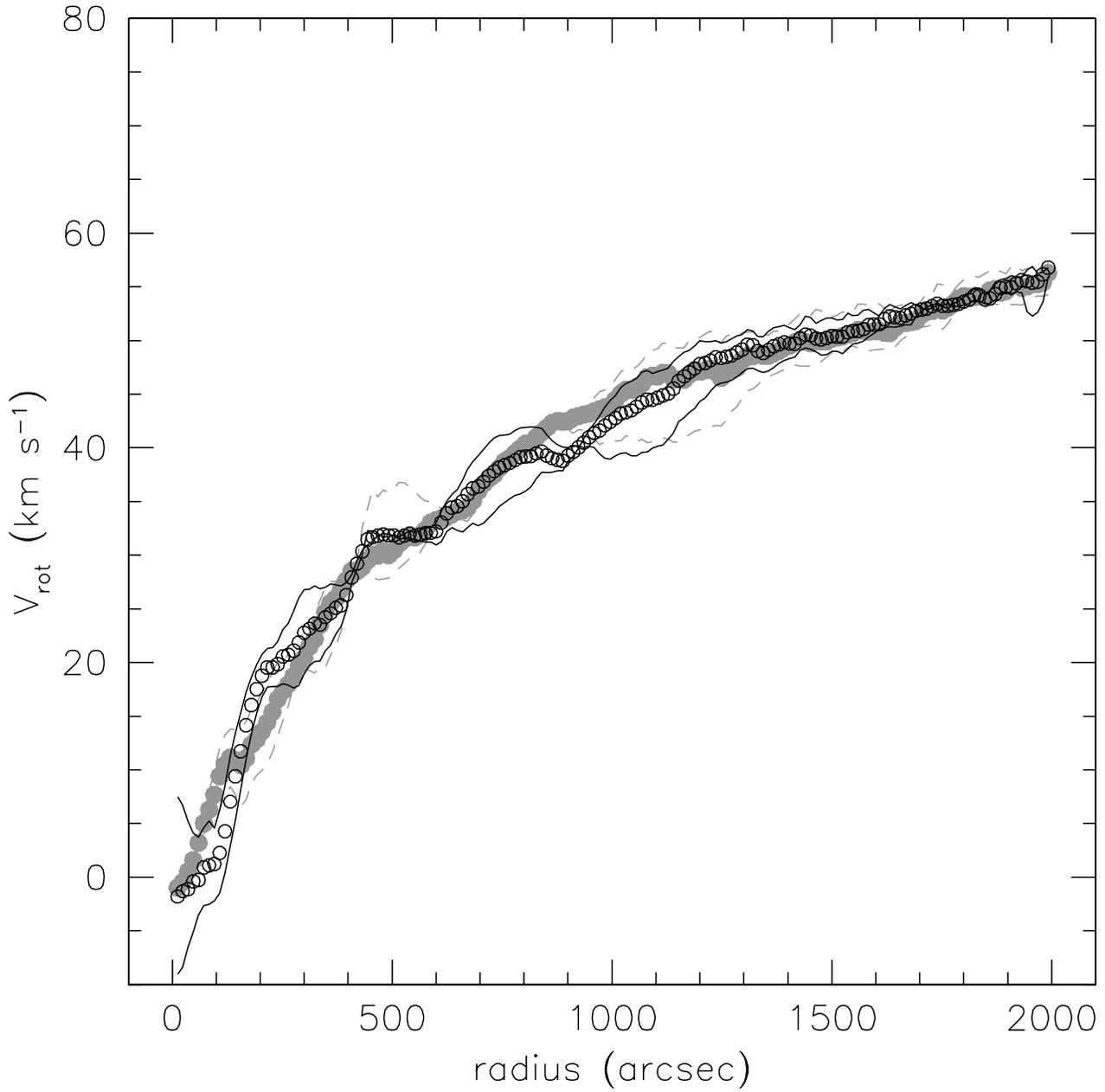,width=\hsize}
\caption{Comparison of the B12 rotation curves derived using the true dynamical center as central position (grey curves), and using the $K$-band 
optical center as central position (black curves). The thin lines represent the curves from the appraching and receding sides separately.
\label{optcenter}}
\end{figure*}

Though the dynamical center is \emph{by definition} the zero-point of
the rotation curve, it is instructive to re-derive the rotation curve
under the assumption that the optical center is the true center of the
galaxy. This is implicitly assumed in many emission line observations
of rotation curves where the optical center of the galaxy is
frequently used to line up the slit of the spectrograph.  We
re-derived the B12 rotation curve fixing the position of the center to
that of the optical center. We kept the run of position angle and
inclination identical to that in the original
model. Fig.~\ref{optcenter} over-plots the two curves. The overall
shape of the curve is insensitive to the precise position of the
center. The small-scale differences are of the same order of magnitude
as the differences between the approaching and receding sides of the
original curve.  One should however not exaggerate the importance of
this offset: had NGC 6822 been at a more typical distance of a few
tens of Mpc, then the angular size of the offset would only be $\sim
1''$ and would have been unnoticed\footnote{ 
As an aside, we refer ahead to the discussion on central mass-density slopes in
Sect.~\ref{sec:innerslope}, where we determine the power-law slope
$\alpha$ of the mass-density profile $\rho \sim r^{\alpha}$ within a
radius of 0.8 kpc. For the B12 case we there find a slope of $\alpha = -0.04
\pm 0.09$. For the curve presented here, determined using the optical center and also using an outer radius of 0.8 kpc,
a similar analysis gives $\alpha = -0.23 \pm 0.24$. The uncertainties are
larger, but the value is consistent with a soft core in NGC~6822. 
The exact position of the dynamical center does not significantly influence the results.}.

\section{Mass components}

\subsection{Stellar component\label{sec:opt}}

We used a mosaic of full-resolution 2MASS $K_s$ images to derive the
surface brightness profile of NGC 6822. The orientation parameters and
the centers of the tilted ring fits to the velocity fields were used
to measure the average surface brightness at each radius.  Prior to
the ellipse integration bright stars were masked out.  The average
surface brightness contribution of remaining foreground stars in the
field away from NGC 6822 was determined and subtracted from the
profile.

The $K_s$ surface brightness of NGC 6822 is low and at large radii the
scatter is substantial. At radii beyond $750''$ the signal of the
galaxy can no longer be distinguished from the sky background.
Fig.~\ref{kbandprof} shows the azimuthally averaged surface brightness
profile, corrected for inclination, determined using the B08 tilted
ring parameters.  The surface brightness was corrected for Galactic
foreground extinction, though this was only a small correction of
0.088 mag in the $K$-band. We assumed NGC 6822 to be optically thin in
$K$, and only applied a geometric inclination surface brightness
correction.

The profile can be well described by an exponential disk. As the outer
parts of the profile could be marginally affected by small
uncertainties in the sky background we have fitted an exponential disk
to the profile at $R<400''$ and find an exponential scale length of
286$''$ or 0.68 kpc.  The central surface brightness is $\mu_0^i(K_s)
= 19.8$ mag arcsec$^{-2}$.  The total absolute magnitude we derive
assuming an exponential disk and extrapolating to infinity is $M_K =
-17.9$ or $L_K = 3.8 \times 10^8\ L_{\odot}$. 

For each resolution we determine the surface brightness profiles using
the relevant tilted ring parameters determined.  At $R>240''$ we
replace the surface brightness profiles with that of the exponential
disk fit as indicated in Fig.~\ref{kbandprof}.  The rotation curve of
the stellar disk was then computed from the extended $K$-band profile
following \citet{cas93} and
\citet{beeg_phd}. The disk was assumed to have a vertical sech$^2$ distribution
with a scale height $z_0 = h/6$ \citep{kruit81}.  The rotation curves
of the stellar component were sampled at the same radii as the
rotation curves. We assume \MLstar\ is constant with radius. While one
expects some modest variation in $\MLstar$ with radius in optical
bands \citep{dejong_grad}, the colour gradients in dwarf LSB galaxies
in the $K$ band tend to be small, so this effect is not likely to be
significant (see also Sect.~\ref{sec:mlstar}).

\subsection{Gas component}

\begin{figure*}
\psfig{figure=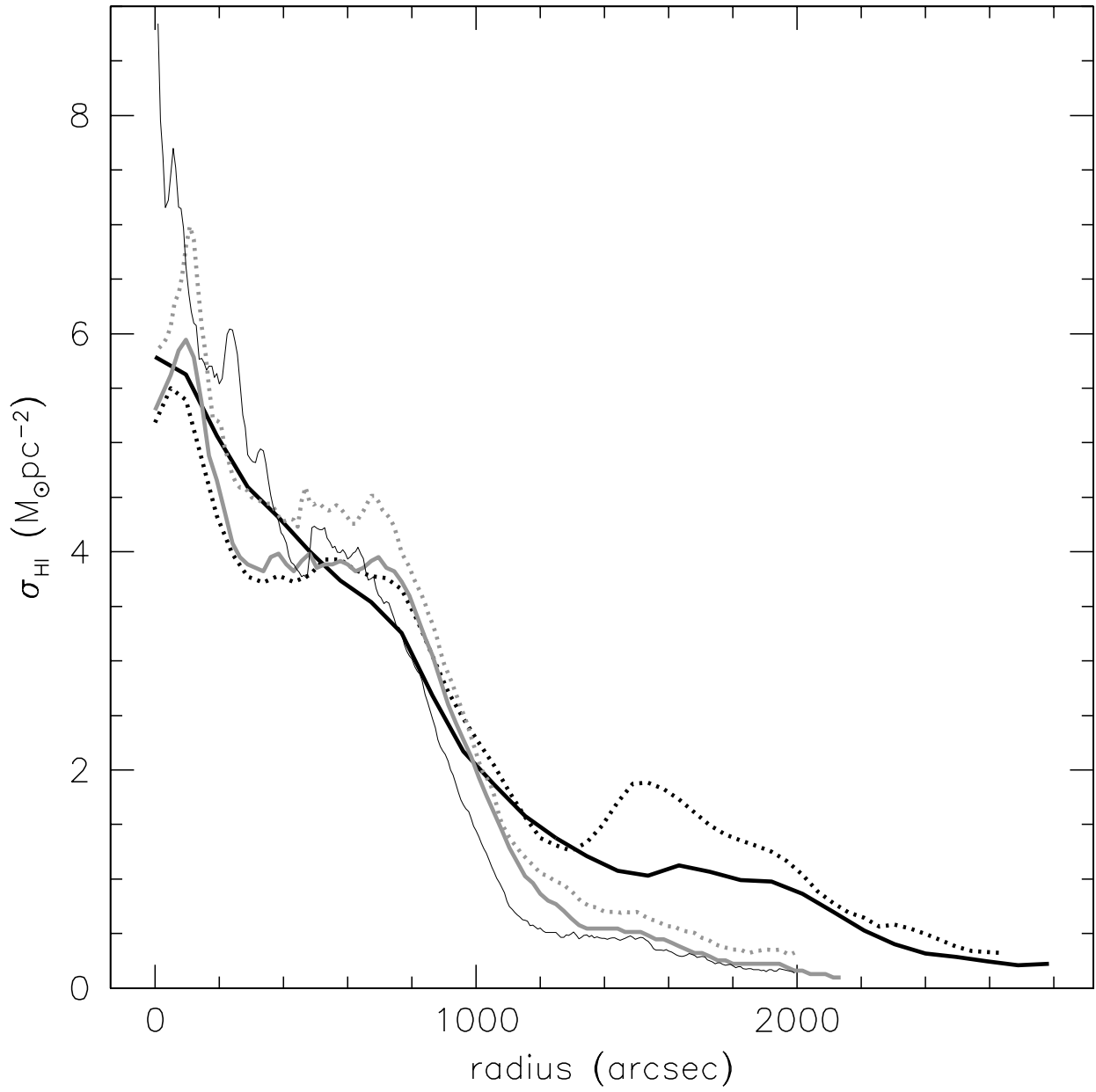,width=\hsize}
\caption{\HI surface density profiles, corrected for inclination.
The various resolutions are distinghuished as follows: B96: full thick 
black line; B48: dotted thick black line; B24: full grey thick line; B12: dotted grey thick line; B08: full thin black line.
\label{hiprof}}
\end{figure*}

\begin{figure*}
\psfig{figure=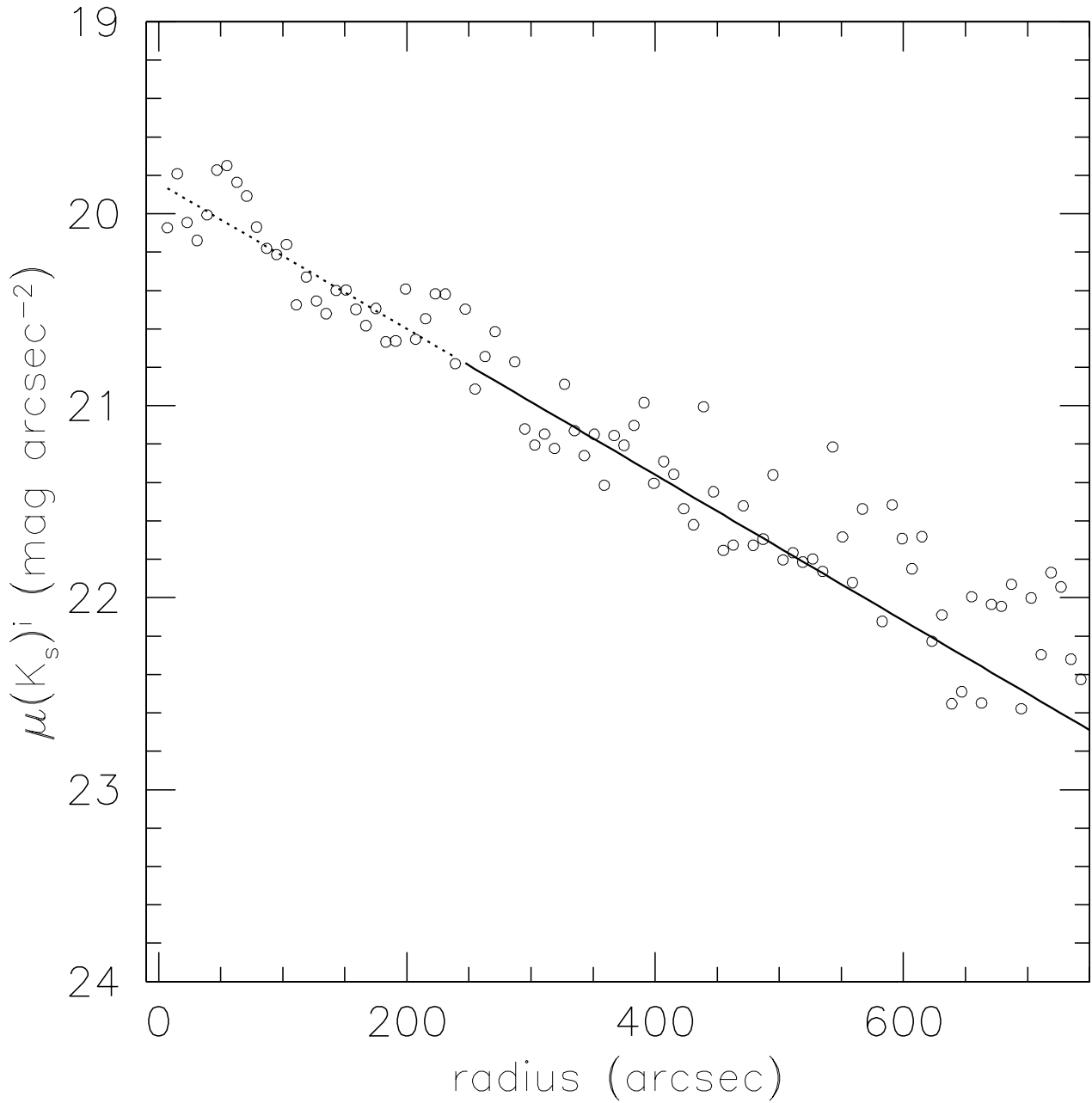,width=\hsize}
\caption{Azimuthally averaged $K$-band surface brightness profile.
The profile is corrected for inclination and Galactic foreground
extinction. Spacing between the points is $8''$. The line indicates
the exponential disk fit used at radii $R>240''$.
\label{kbandprof}}
\end{figure*}

We used the orientation parameters of the tilted ring fits to
determine the azimuthally average \HI surface density profiles using
the total column density maps for each resolution.  The \HI surface
density profiles are presented in Fig.~\ref{hiprof}.  In deriving the
corresponding rotation curve with the {\sc gipsy} task {\sc rotmod} we
scaled the surface density profiles by a factor of 1.4 to take the
contribution of helium and metals into account.

We assume that the gas is distributed in an infinitely thin disk.
This assumption is not crucial. Though one might expect dwarf galaxies
to have a more puffed up disk than a typical spiral galaxy, the shape
and amplitude of the gas rotation curve depends only very mildly on
the thickness of the disk.

To use the B12 curve as an example, for the infinitely thin gas disk
case the maximum rotation velocity of the gas component is
$17$~\kms. When increasing the thickness of the disk, this maximum
rotation velocity drops only very slowly: for an exponential scale
height $z=0.5$ kpc, it still measures $\sim 14.5$~\kms. This value for
$z$ is already approaching the value of the radial scale length of the
stellar distribution and likely already too large. We thus adopt the
thin disk approximation.

\subsection{Halo models}
The dark halo component differs from the previous two in that we are
interested in parametrising this component assuming some fiducial
model. The choice of this model is the crux of most of the DM analyses
in the literature, and many models exist.  These can all be broadly
distinguished in two groups: halo models with a core, and halo models
with a cusp. An example of the first category is the pseudo-isothermal
halo, an example of the latter the CDM NFW halo.

\subsubsection{Pseudo-isothermal halo}

The spherical pseudo-isothermal (ISO) halo has a density profile
\begin{equation} 
\rho_{ISO}(R) = \frac{\rho_0}{1 + ({{R}/{R_C}})^2 }, 
\end{equation} 
where $\rho_0$ is
the central density of the halo, and $R_C$ the core radius of the
halo.  The corresponding rotation curve is given by
\begin{equation} V(R) = \sqrt{ 4\pi G\rho_0 R_C^2 \Bigl[ 1 -
{{R_C}\over{R}}\arctan \Bigl( {{R}\over{R_C}} \Bigr) \Bigr] }.
\end{equation}
The asymptotic velocity of the halo, $V_{\infty}$, is given by
\begin{equation} 
  V_{\infty} = \sqrt{ 4 \pi G \rho_0 R_C^2 }.
\end{equation}
To characterise this halo only two of the three parameters
$(\rho_0, R_C, V_{\infty})$ are needed, as equation (3) determines the
value of the third parameter.  

\subsubsection{NFW halo}

The NFW mass density distribution takes the form \citep{NFW96}
\begin{equation}
\rho_{NFW}(R) = \frac{\rho_0}{(R/r_0)[1+(R/r_0)]^{2}}
\end{equation}
%\rho_{NFW}(R) = \frac{\rho_0}{\left(R/R_s\right)
%\left(1+ R/R_s\right)^2}
%\end{equation}
where $r_0$ is the characteristic radius of the halo and
$\rho_0$ the characteristic density \citep{NFW96,blais00}.
This mass distribution gives rise to a halo rotation curve
\begin{equation}
V(R) = V_{200} \left[\frac{\ln(1+cx)-cx/(1+cx)}
{x[\ln(1+c)-c/(1+c)]}\right]^{1/2},
\end{equation}
where $x = R/R_{200}$.
It is characterised by a concentration parameter $c =
R_{200}/R_s$ and a radius $R_{200}$. These are directly related to
$R_s$ and $\rho_i$, but are used instead as they are a convenient way
to parametrise the rotation curve.  The radius $R_{200}$ is the radius
where the density contrast exceeds 200, roughly the virial radius
\citep{NFW96}. The characteristic velocity $V_{200}$ of the
halo is defined in the same way as $R_{200}$.  These parameters are
not independent and are set by the cosmology.

\subsection{Stellar Mass-to-light ratios and weighting\label{sec:mlstar}}

One of largest uncertainties in any mass model is the value of the
stellar mass-to-light ratio
\MLstar. Though broad trends in \MLstar\  have been measured and modelled 
\citep[e.g.][]{botje97,bell_pops}, the precise value for an individual galaxy
is not well known, and depends on extinction, star formation history,
Initial Mass Function, etc.  The value of \MLstar\ cannot be
constrained using the rotation curve alone \citep{maxdisk86,lake89}
and some assumptions must be made.  We therefore present disk-halo
decompositions using four different assumptions.

{\bf Minimum disk.} This model assumes that the observed rotation
curve is due entirely to DM. This gives an upper limit on how
concentrated the dark mass component can actually be.  It is not a
realistic model \emph{per se}, because it ignores the gas disk which
is obviously present. However, we present it here for ease of
comparison with cosmological simulations, a large fraction of which
tend to only model the dark matter component.

{\bf Minimum disk + gas.}  The contribution of the atomic gas (H{\sc
i} and He) is taken into account, but \MLstar\ is assumed to be
zero. 

{\bf Constant \MLstar.} Here we choose a value for \MLstar\
appropriate for NGC 6822 based on its stellar content, colours and likely
star formation history.  Unfortunately reliable integrated colours for
NGC 6822 are rare due to its large angular size.  We can however use
indirect arguments to arrive at a value for \MLstar.  Furthermore we
are interested here only in the $K$-band value which is rather
insensitive to effects of extinction and recent star formation.

\citet{marc_phd97} presents $K'$-band rotation curve fits of a large
sample of HSB and LSB galaxies in the Ursa Major cluster.  We consider
the rotation curve fits derived using the Bottema disk prescription
\citep{botje97}. This recipe states that the maximum rotation velocity
of the disk is 63\% of the maximum rotation velocity observed (and is
derived from measurements of stellar velocity dispersions). The
distribution of $\MLstar{}_K$ ratios derived with the Bottema disk for
the dwarf and LSB galaxies in the UMa sample peaks at $\MLstar{}_K =
0.4 \pm 0.05$.
If the colours of NGC 6822 are comparable to those found for other dwarfs
and LSB galaxies (namely $B-V \sim 0.5$, $B-R\sim 0.8$ and $V-I \sim
0.8$; \citealt{edb_phot95}) we can use Table 3 in \citet{bell_pops}
to derive an approximate value for
\MLstar. For their formation epoch model with burst we find values 
for $\MLstar{}_K$ between 0.35 and 0.40. Similar values are found for
the infall models and the hierarchical models. 
From detailed population synthesis modelling of NGC 6822,
\citet{gallart2} find that the total mass of stars and stellar remnant
ever formed in NGC 6822 must be $\sim 9 \cdot 10^7$ \Msun. Combined with
the $K$-band luminosity derived earlier, this gives a value
$\MLstar{}_K = 0.24$.

Taking the above into account we adopt a value of $\MLstar{}_K = 0.35$
as representative of the stellar population in NGC 6822.

{\bf Maximum disk.} The rotation curve of the stellar component is
scaled to the maximum value allowed by the observed rotation curve,
but with the restriction that the DM density gradient is required to
be flat or negative at all radii (thus avoiding a  ``hollow
halo'') \citep{maxdisk86}.  

Each of the rotation curves was fitted using the {\sc gipsy} task {\sc
rotmas}. The program determines the best-fitting combination of $R_C$
and $V_{\infty}$ (for the pseudo-isothermal halo) or $c$ and $V_{200}$
(for the NFW halo), using a least squares fitting routine.  We
assigned weights to the data points inversely proportional to the
square of their uncertainty.

\section{Results}

\begin{figure*}
\hbox{\psfig{figure=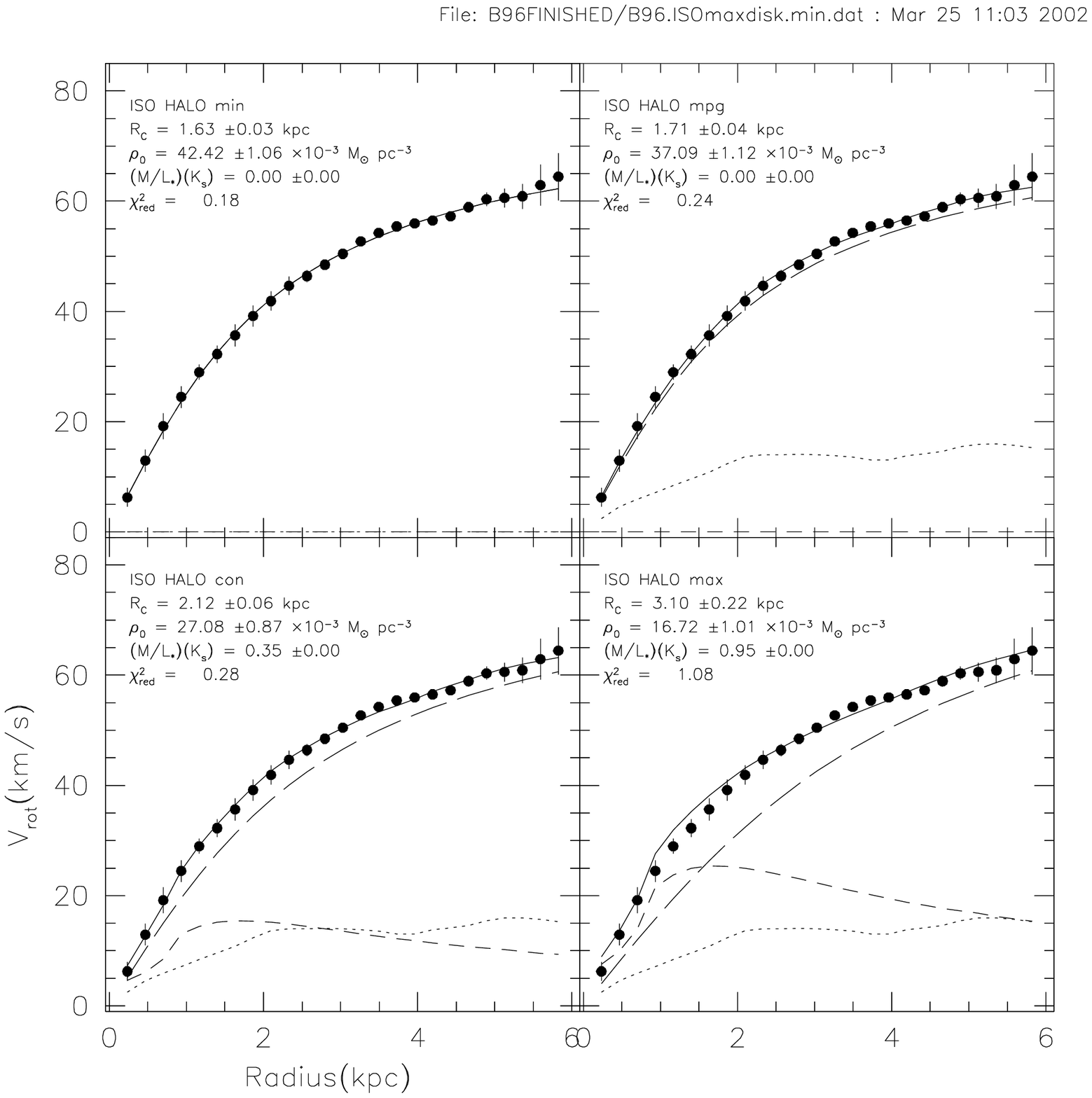,width=0.45\hsize,clip=true}\psfig{figure=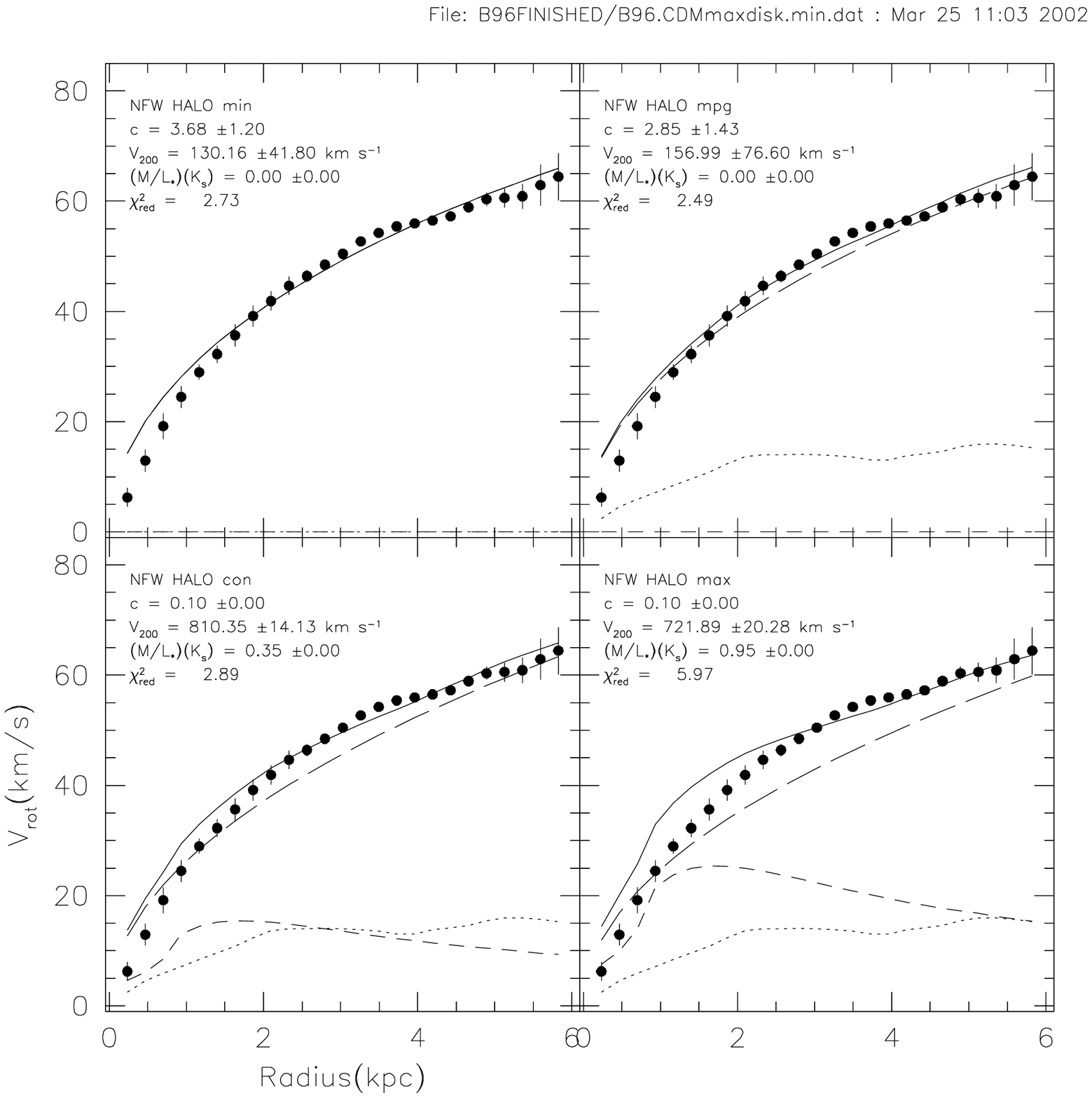,width=0.45\hsize,clip=true}}

%\hbox{\psfig{figure=B96.innerISO.panel.pap.ps,width=0.45\hsize,clip=true}\psfig{figure=B96.innerCDM.panel.pap.ps,width=0.45\hsize}}

\caption{B96 mass models. The left column shows the fits made using a 
pseudo-isothermal (ISO) halo. The right column shows the NFW halo
fits. The top row shows fits made to the entire rotation curve, while
the bottom row shows fits to the inner curve $(R<1000'')$.  In each
panel the following fits are shows: minimum disk (top-left), minimum
disk+gas (top-right), constant \MLstar\ (bottom-left), and maximum
disk (bottom-right). Fit parameters are given in the sub-panels.  An
error of ``0.00'' indicates that this parameter was fixed during the
fitting.
\label{curves96}}
\end{figure*}

\begin{figure*}
\hbox{\psfig{figure=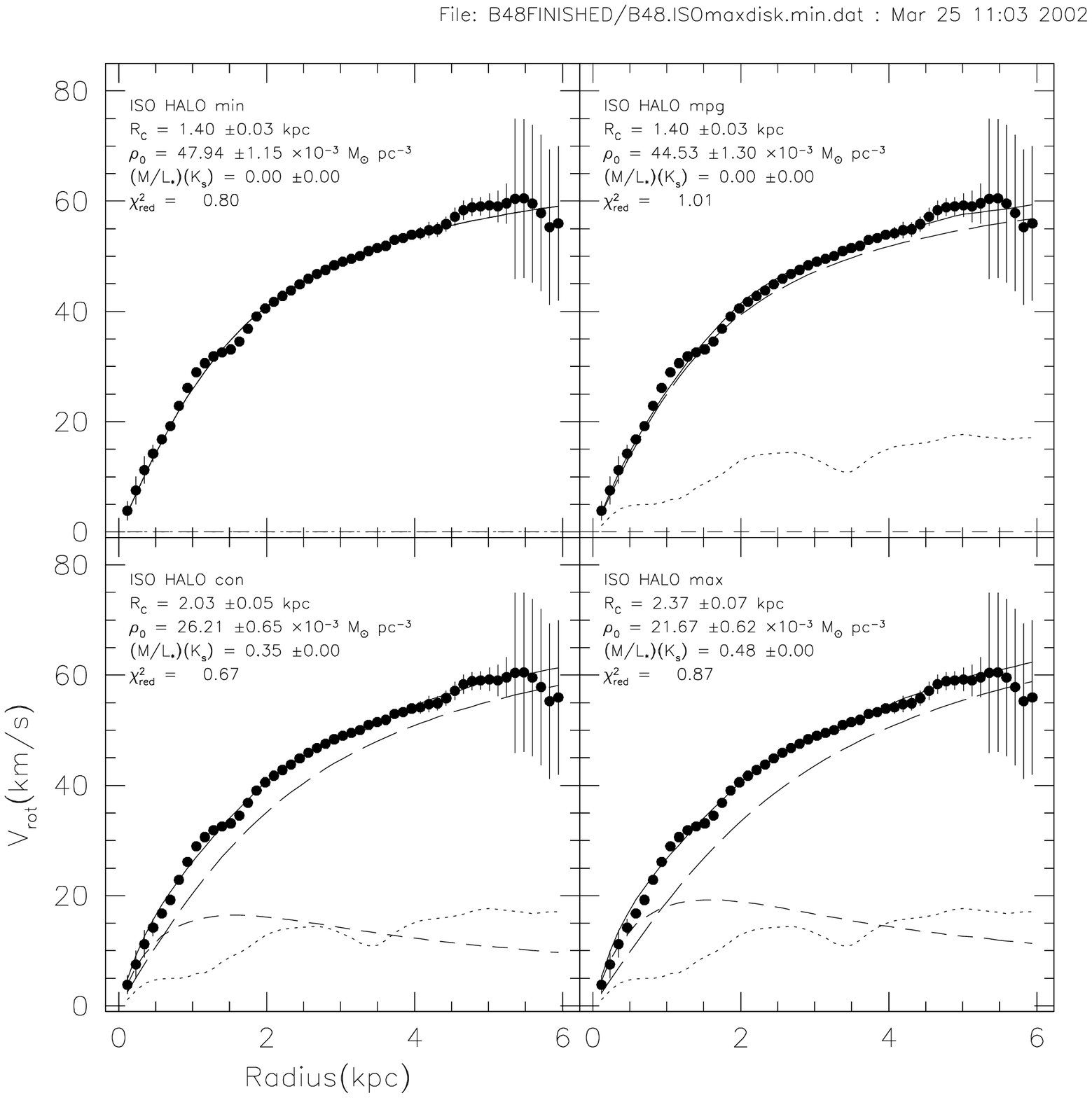,width=0.45\hsize,clip=true}\psfig{figure=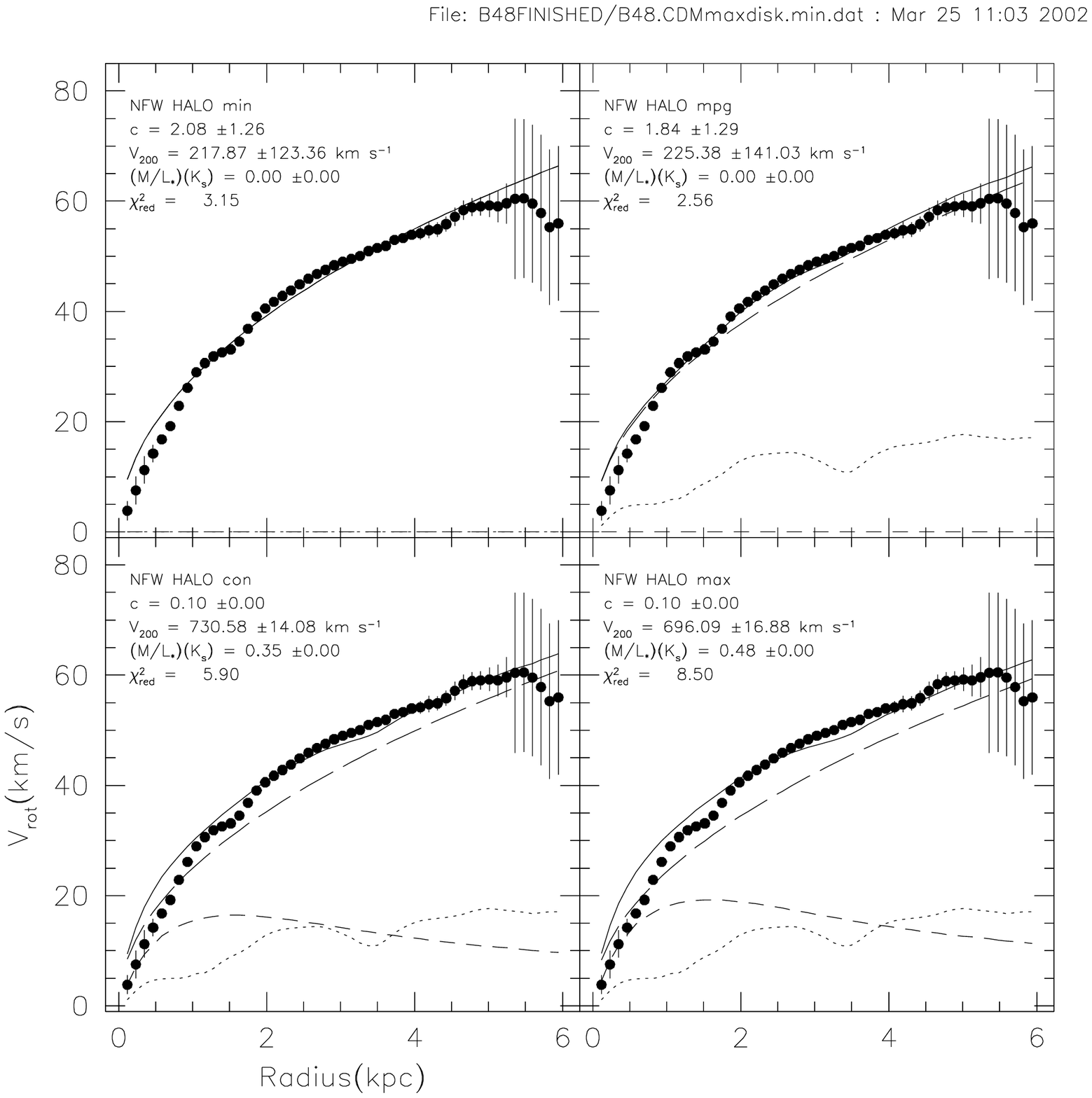,width=0.45\hsize,clip=true}}

%\hbox{\psfig{figure=B48.innerISO.panel.pap.ps,width=0.45\hsize,clip=true}\psfig{figure=B48.innerCDM.panel.pap.ps,width=0.45\hsize,clip=true}}
\caption{B48 mass models. See Fig.~\ref{curves96}.\label{curves48}}
\end{figure*}

\begin{figure*}
\hbox{\psfig{figure=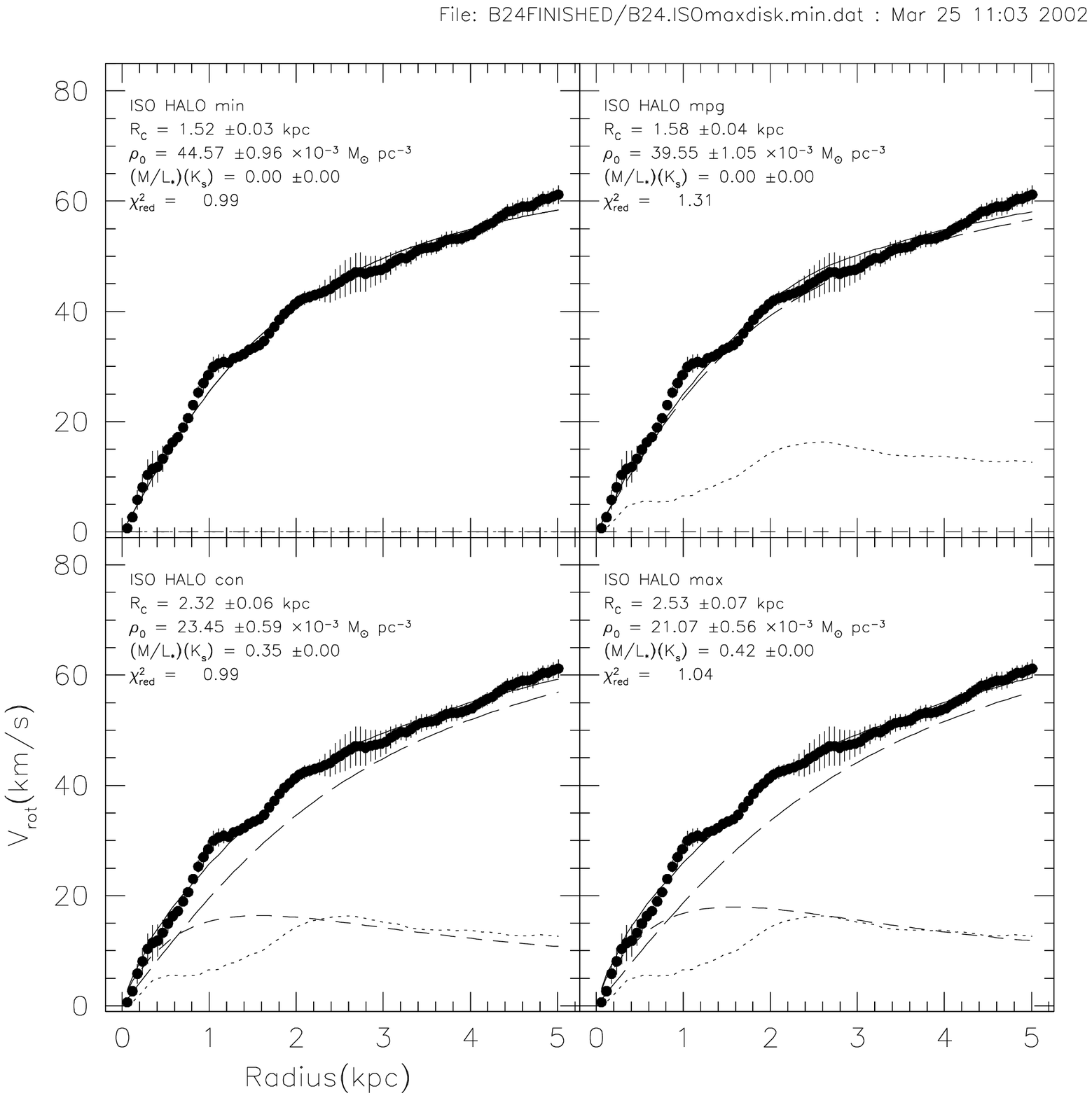,width=0.45\hsize,clip=true}\psfig{figure=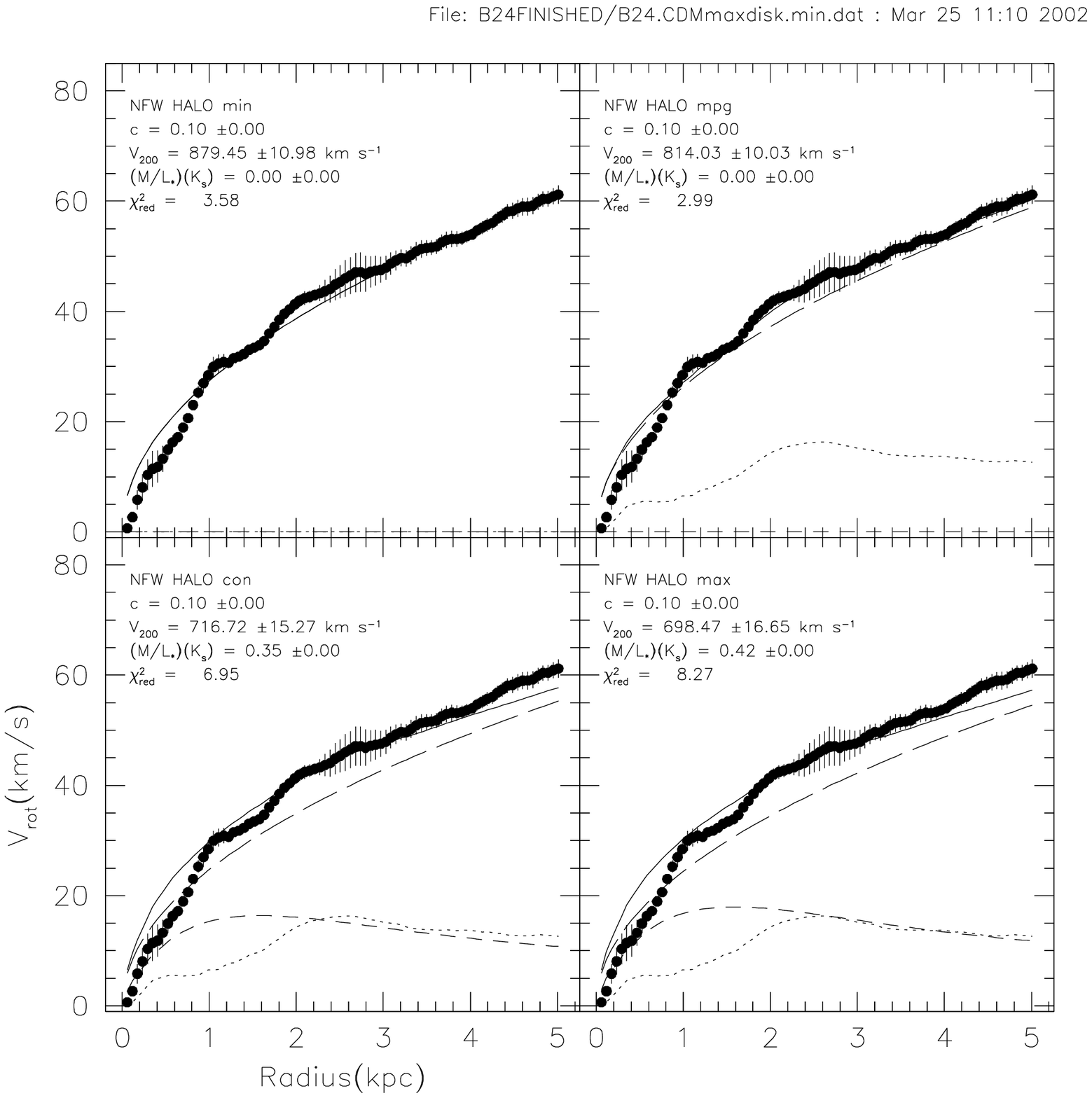,width=0.45\hsize,clip=true}}

%\hbox{\psfig{figure=B24.innerISO.panel.pap.ps,width=0.45\hsize,clip=true}\psfig{figure=B24.innerCDM.panel.pap.ps,width=0.45\hsize,clip=true}}
\caption{B24 mass models. See Fig.~\ref{curves96}.\label{curves24}}
\end{figure*}

\begin{figure*}
\hbox{\psfig{figure=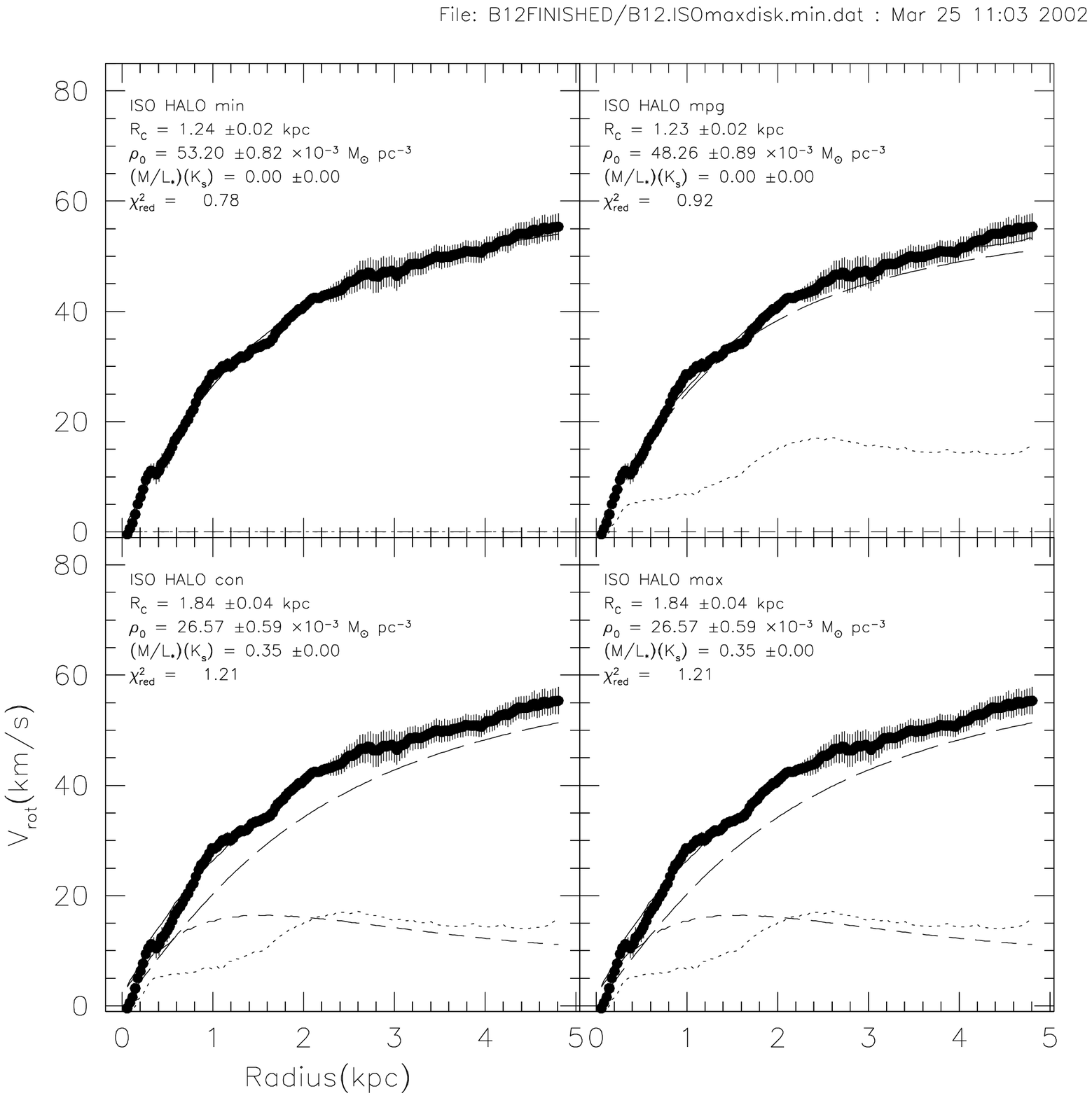,width=0.45\hsize,clip=true}\psfig{figure=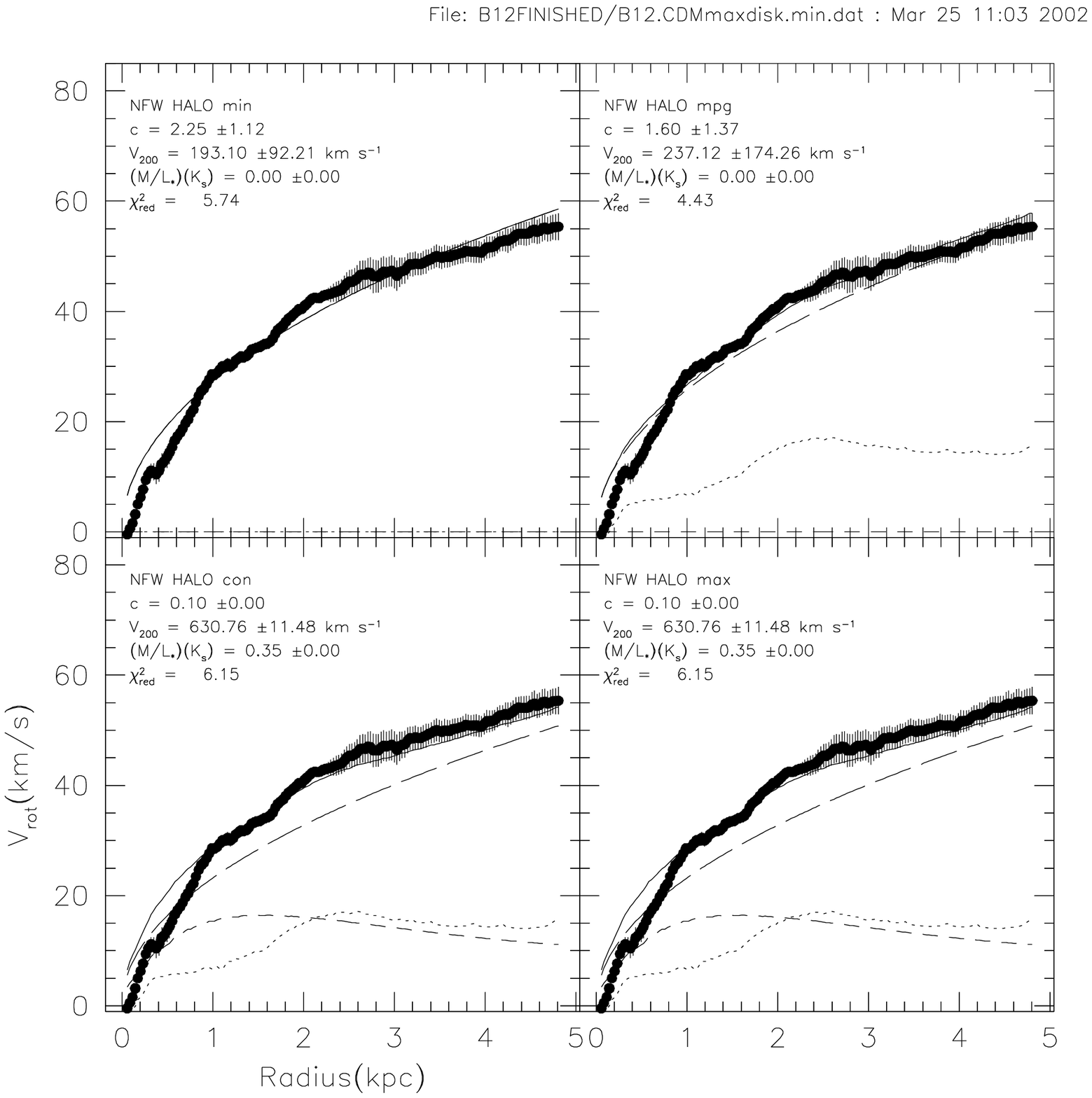,width=0.45\hsize,clip=true}}

%\hbox{\psfig{figure=B12.innerISO.panel.pap.ps,width=0.45\hsize}\psfig{figure=B12.innerCDM.panel.pap.ps,width=0.45\hsize,clip=true}}
\caption{B12 mass models. See Fig.~\ref{curves96}.\label{curves12}}
\end{figure*}

\begin{figure*}
\hbox{\psfig{figure=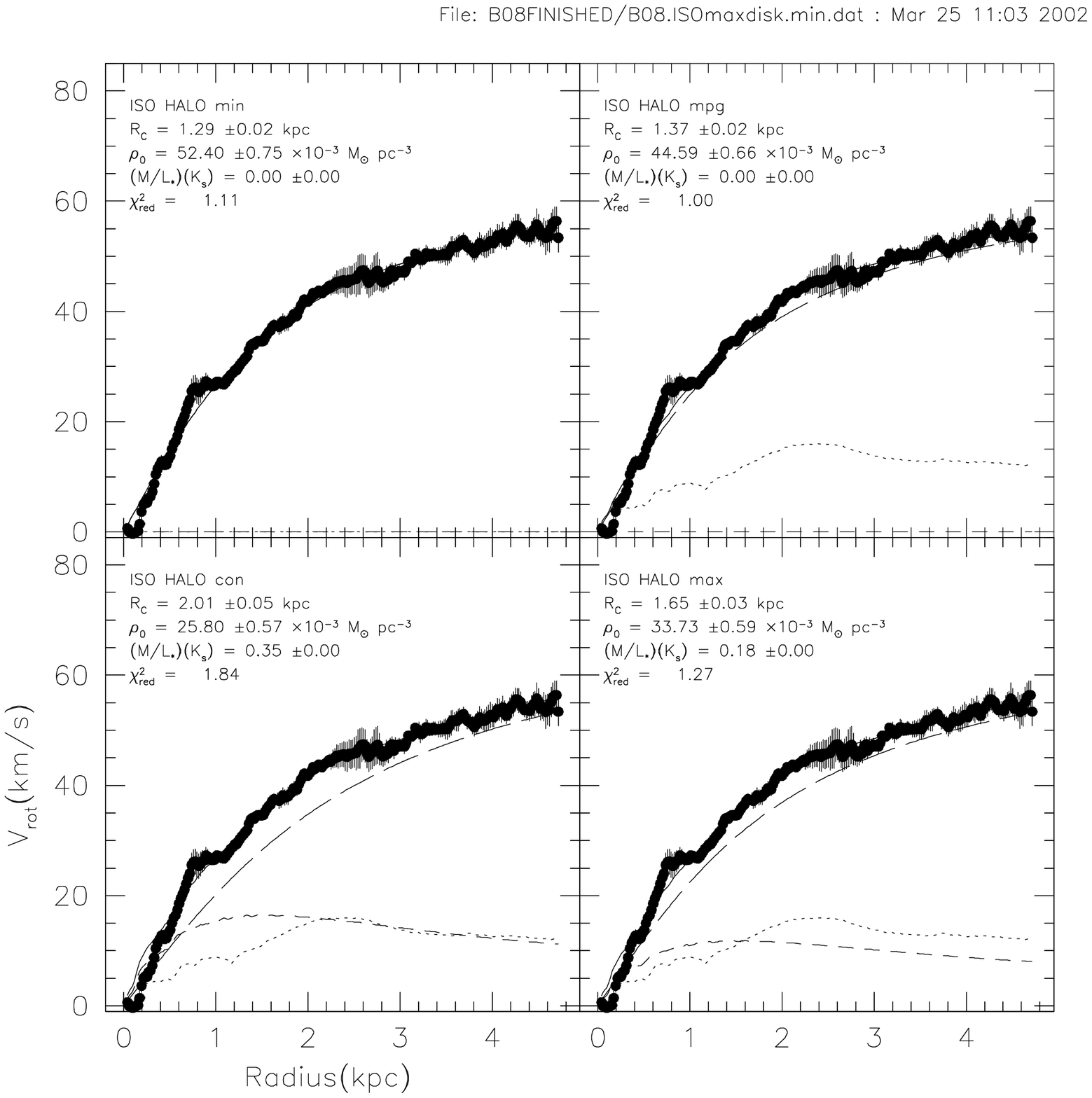,width=0.45\hsize,clip=true}\psfig{figure=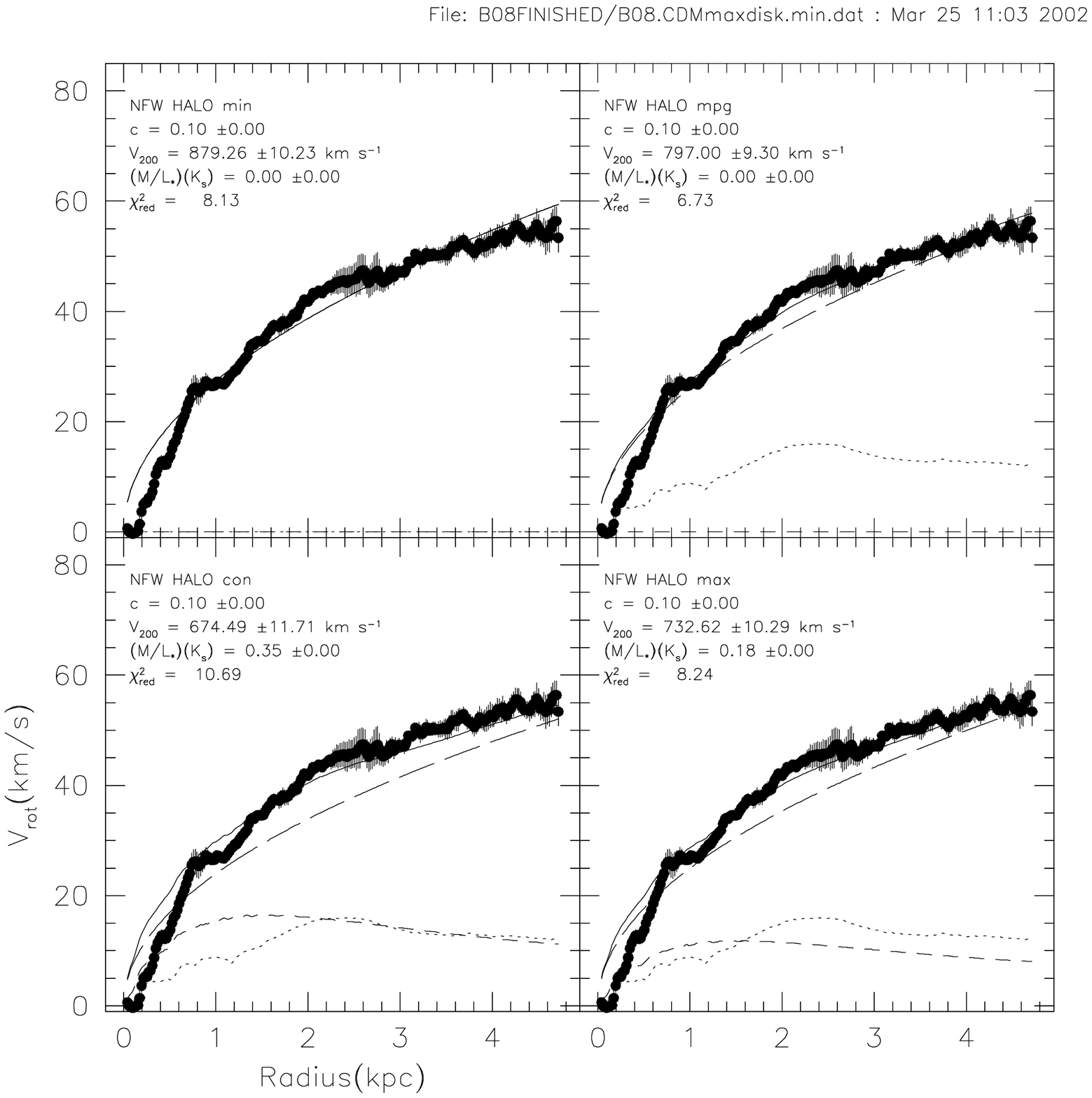,width=0.45\hsize,clip=true}}

%\hbox{\psfig{figure=B08.innerISO.panel.pap.ps,width=0.45\hsize}\psfig{figure=B08.innerCDM.panel.pap.ps,width=0.45\hsize}}
\caption{B08 mass models. See Fig.~\ref{curves96}.\label{curves08}}
\end{figure*}

Figures \ref{curves96}-\ref{curves08} show the rotation curves at all
resolutions plotted for the four assumptions on \MLstar\ for both NFW
and ISO halo models.  To rule out the possibility that the outer parts
of NGC 6822, which might be affected by tidal effects, affect our
conclusions we also present separate fits to just the inner part of
the galaxy out to to $R=1000''$ (2.4 kpc) or the edge of the main
inner disk.  Tables~\ref{nfwfits} and \ref{isofits} present the fit
parameters for all models presented here.

\begin{figure*}
\psfig{figure=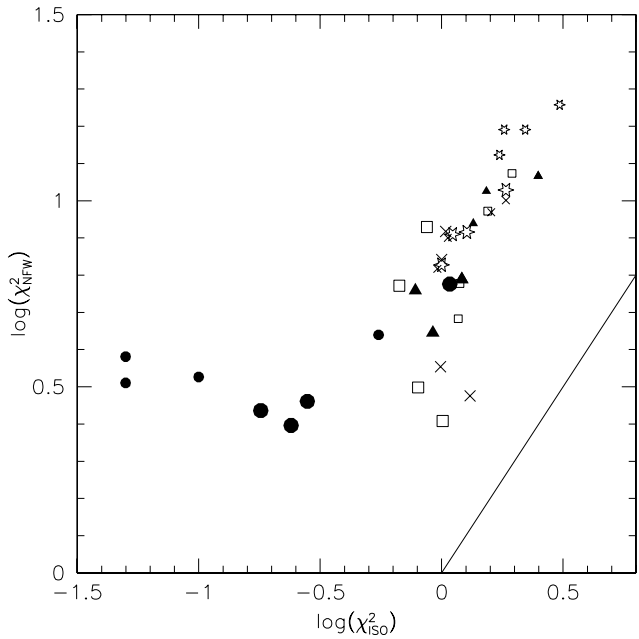,width=0.5\hsize}
\caption{Reduced chi2 plot, $\chi^2(ISO)< \chi^2(NFW)$ at all resolutions and \MLstar. The line is the line of equality. Large symbols represent the fits to the complete curves, small symbols fits to the inner curves. 
Filled circles: B96; open squares: B48; crosses: B24; filled triangles: B12; and stars: B08.
\label{chi2}}
\end{figure*}

An inspection of Tables \ref{nfwfits} and \ref{isofits} shows that the
goodness-of-fit is in all cases better for ISO than for NFW. The only
exceptions are some of the maximum disk fits where both ISO and NFW
are difficult to fit, as NGC 6822 is not a maximum disk galaxy. A
comparison between the $\chi^2$ values is shown in Fig.~\ref{chi2}. 

The maximum disk values of \MLstar\ generally have an uncertainty of
$\pm \sim 0.1$. Changing the values within this range has no
discernible effect on the quality of the fits.  One point of note is
that for the B08 case the maximum disk value for $\MLstar$ is less
than that for the constant \MLstar\ case. The B08 constant \MLstar\
fit thus results in a hollow halo.

Many of the NFW fits have fit parameters that do not make physical
sense.  The large majority of the fits prefer $c \leq 0$ and $V_{200}
\rightarrow \infty$, which is another way of saying that the fit 
procedure is trying to fit a $V\sim R^{1/2}$ curve to a $V \sim R$
curve by stretching it to infinity, thus trying to take out the
curvature in the model.  Where this happened we fixed the $c$
parameter to $c=0.1$ with only $V_{200}$ as the free parameter.  Such
a value is still outside the range predicted by cosmological
simulations where values of $c\sim 10$ would be expected.
There is a general trend for the value of the maximum disk $\MLstar$
to decrease with increasing resolution. It is not clear what the cause
of this effect is. The inner slopes of the rotation curve and stellar
curve change subtly with resolution, but not enough to be the sole
cause. It is likely that small differences between the tilted ring
models also play a role.  

We note that at all resolutions we find that (restricting ourselves to
the ISO model for a moment) the min+gas model usually has the smallest
value of $\chi^2$, and that for all cases $\chi^2_{m+g} <
\chi^2_{con}$.   We have tried to constrain the value of $\MLstar$ by
making a ``best fit'' to the rotation curves using ISO models: i.e.\
we also let $\MLstar$ be a free parameter in the fits.  Unfortunately
the different resolutions make it difficult to constrain $\MLstar$ in
this manner. The B96 data is best fitted with a stellar component with
$\MLstar{}_K = 0.10 \pm 0.13$, while the B48 and B24 curves both demand
$\MLstar{}_K = 0.31 \pm 0.04$. For the B08 and B12 data such a ``best fit'' gives
unphysical results (negative $\MLstar$). A constrained fit with
$\MLstar \geq 0$ yields $\MLstar = 0$ (i.e.\ the minimum disk+gas case)
as best fit for these resolutions. Though it is clear that $\MLstar$ is difficult to constrain using only rotation curve information, the results do suggest
that 
the stellar population is dynamically
unimportant.

\section{Discussion}
\subsection{Inner slopes\label{sec:innerslope}}
With the high-resolution rotation curves we can determine the inner
slope of the mass-density distribution. The rationale behind this is
described in \citet{paper3}.  In summary, the various halo models make
specific predictions about the slope of the dark matter mass density
distribution in the inner parts.  If we approximate this distribution
with a power-law $\rho \sim r^{\alpha}$, then the ISO halo predicts
$\alpha = 0$, while the NFW halo predicts $\alpha=-1$.  In
\citet{paper3} the inner mass density slopes of a large sample of LSB
galaxies was derived, and the most representative value was $\alpha =
-0.2 \pm 0.2$, i.e.\ closer to pseudo-isothermal than to CDM.

\begin{figure*}
\hbox{\psfig{figure=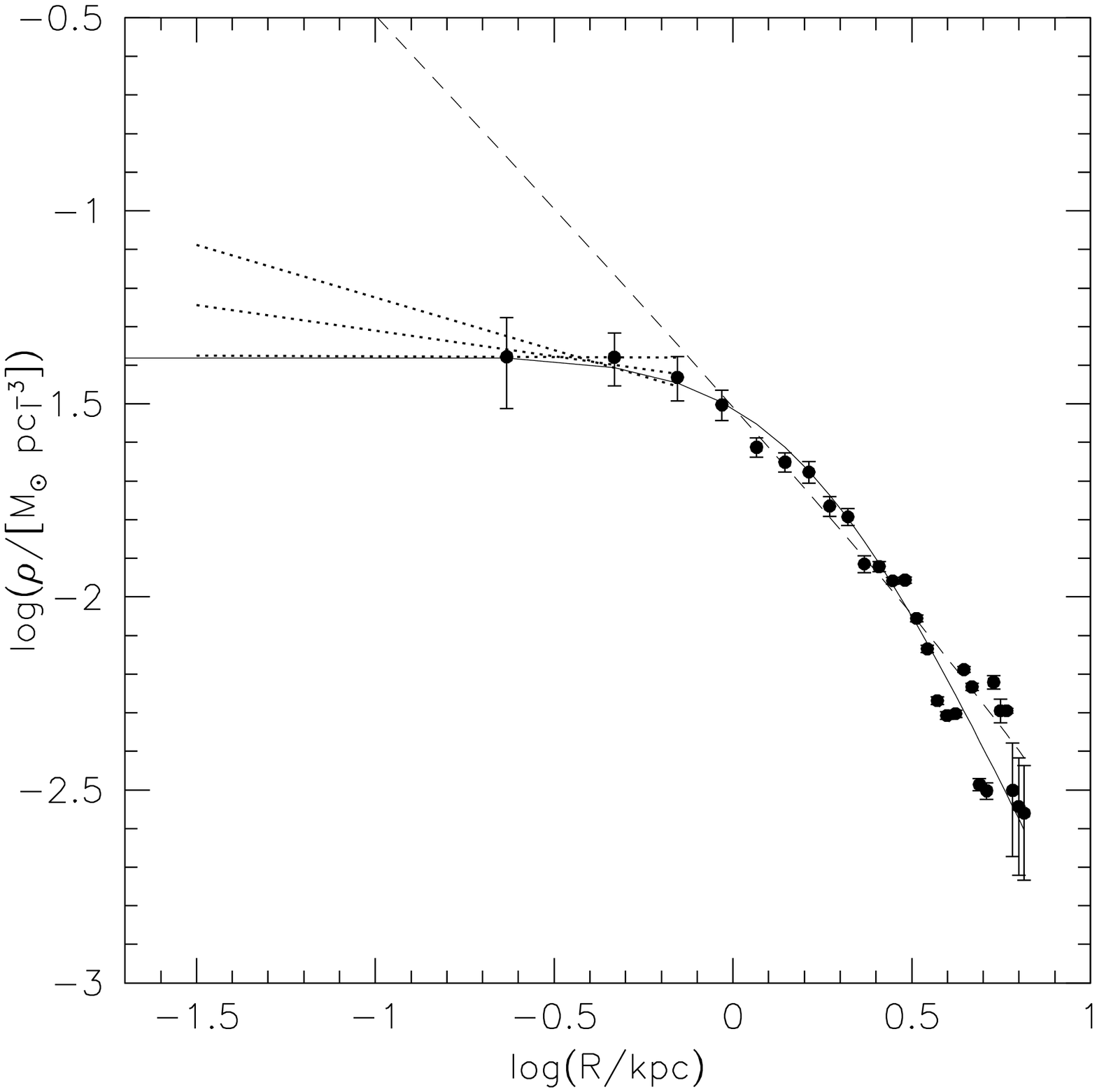,width=0.45\hsize}
\psfig{figure=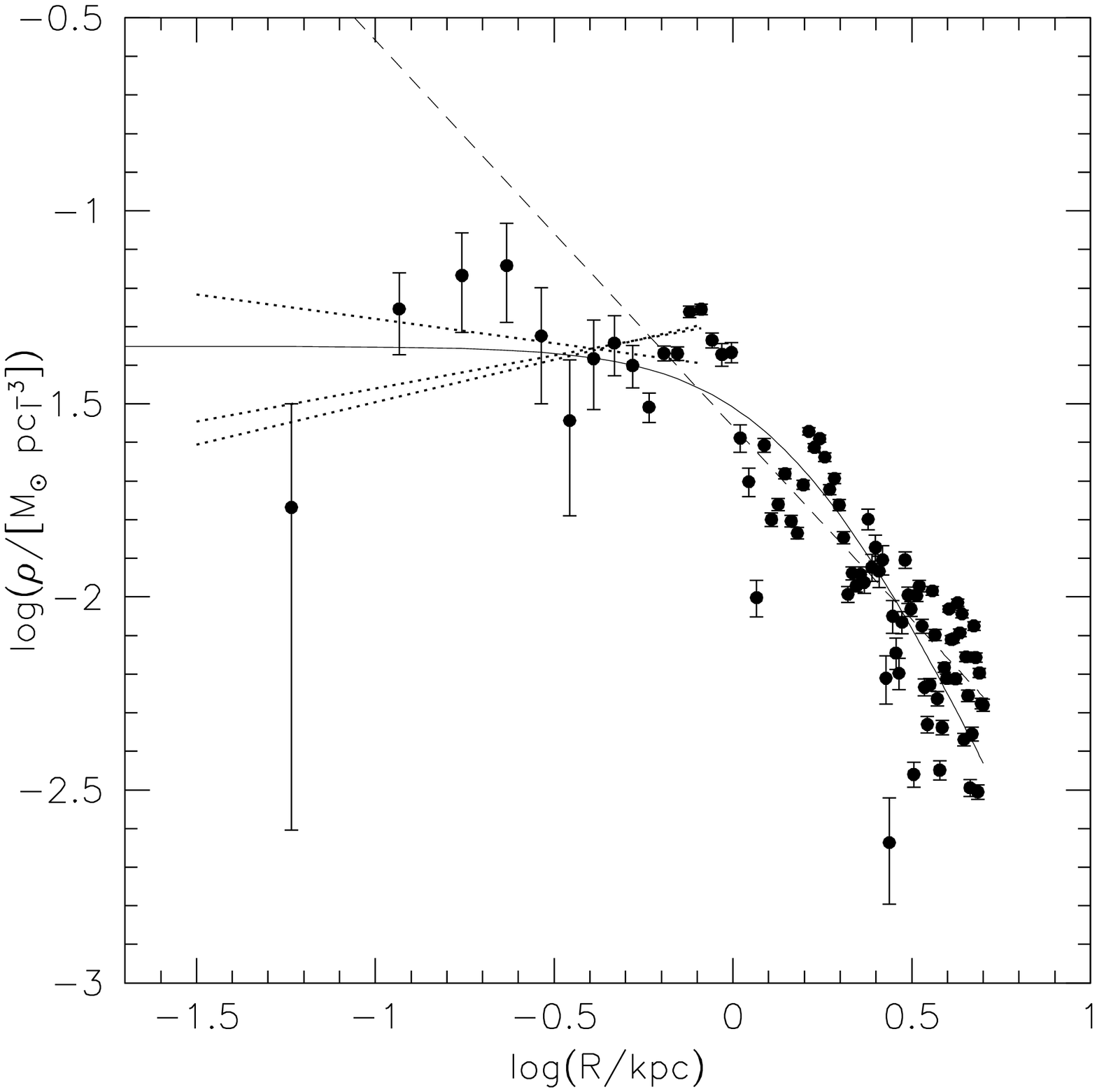,width=0.45\hsize}}
\caption{Mass-density profiles for the B96 (left) and B24 (right) models.
The long-dashed lines shows the best-fitting minimum disk NFW model,
the full line the best-fitting ISO model. The dotted lines represent
the power-law fits to the data at $R<0.8$ kpc.
\label{profiles}}
\end{figure*}

Here we derive the slopes of the mass density distribution of NGC 6822 at
various resolutions using the inversion method described in
\citet{paper3} where we compute the mass density profile and
fit a power-law to the mass-density profile.  Fig.~\ref{profiles} shows two
examples. We show the B96 and B24 mass-density profiles, with the best
fitting NFW and ISO (minimum disk) models over-plotted. The NFW models
have central densities that are discrepant by a factor of 10 or more.

We measure the slope within $R = 0.8$ kpc. This is an arbitrary
choice, motived by the fact that we want to use at least 3 data points
to determine the slope of the B96 curve. Other choices are obviously
possible, and may result in slightly different values of the slope,
but will not affect the conclusions.  The resulting values are given
in Table~\ref{slope}. For the B08 data a meaningful slope could not be
derived due to the large gradients in the rotation velocity that are
found at small scales. The values are at all resolutions significantly
different from the NFW values. Note that there is no systematic
steepening of the slope with resolution, which one would expect if the
cusp was hidden by resolution effects. With a resolution of only tens
of parsecs it is not clear that there is still room for a cusp in
NGC 6822.

\subsection{Dark matter, feedback and cosmology}

\begin{figure*}
\psfig{figure=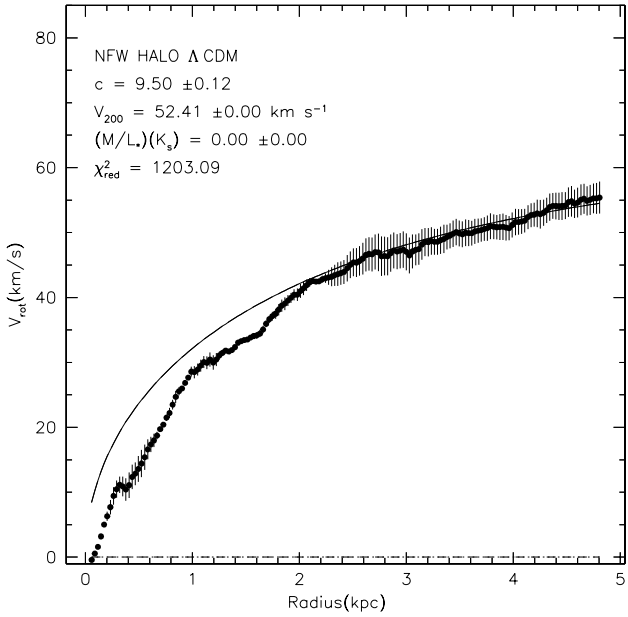,width=0.5\hsize}
\caption{Minimum disk halo for B12 using parameters as required by CDM simulations assuming that $V_{\rm max} = V_{200}$.
\label{cosmohalo}}
\end{figure*}

In the previous sections we have made NFW fits to the rotation curves
disregarding any of the predictions that CDM makes for $c$, $V_{200}$
and their relation. We will compare the observed rotation curve with
cosmological predictions. If we assume that $V_{200}
\simeq V_{\rm max} \simeq 55$ \kms, we find for $\Lambda$CDM that
$c\simeq 9.5$ \citep{NFW97}. The total mass $M_{200}$ of this halo is
$M_{200} = 5.2 \cdot 10^{10}$ \Msun. This predicted halo is
over-plotted on the observed curve in Fig.~\ref{cosmohalo}, and
over-predicts the rotation velocity in the inner parts by a
significant amount.

To look at the problem from a different perspective, let us assume
\MLstar=0.35. This gives a stellar mass $M_*=1.3\times 10^8\ M_{\odot}$.
The total (atomic) gas mass is $1.5 \times 10^8\ M_{\odot}$ (HI+He)
and the total observed baryonic mass is $2.8 \times 10^8\
M_{\odot}$. A universal baryon fraction of $\sim 0.09$ (see
e.g.\ \citealt{white&fabian}) implies a total mass $M_{200} = 3.1
\times 10^9\ M_{\odot}$.  For a NFW halo this implies $V_{200} = 21.5$
\kms\ and $R_{200} = 28$ kpc.  These values are not consistent with
the observed curve, and imply (apart from the inability of the model
to fit the data) that either the baryon fraction in NGC 6822 differs
significantly from the universal value or that large amounts of
baryons have been expelled, as we will discuss below in more detail.

Let us work out the \emph{observed} matter fractions in NGC 6822.
We use the ISO models, and determine the amount of dark matter within
a radius of 5 kpc, the outer edge of the \HI\ disk.  This yields a
total dark mass out to 5 kpc of $3.2\times10^9\ M_{\odot}$.  With an
observed baryonic mass of $2.8 \times 10^8\ M_{\odot}$ we find that
NGC 6822 is heavily dark matter dominated with $M_{vis}/M_{dark}(R<5
\mbox{kpc}) = 0.09$. This is comparable to values found in LSB galaxies
and other dwarf galaxies (de Blok \& McGaugh 1997).  The corresponding
baryon fraction is $f_b = 0.080$, close to the universal baryon
fraction.

The visible matter in NGC 6822 is thus a minor component of the total
galaxy system. This makes it harder to explain the observed
core-dominated dark matter distribution as resulting from a NFW halo
modified by feedback.  The term feedback is often used
indiscriminately to indicate star formation and evolutionary processes
that affect the mass distribution in a galaxy.  In fact, there are two
distinct forms of feedback. One of them is commonly observed in
galaxies, also in NGC 6822, and usually shows itself as small-scale
redistribution of mass in the disk.

The second form of feedback is more catastrophic, mostly theoretical,
and was introduced to explain the discrepancy between observed and
theoretical dark matter distributions. In essence it invokes
large-scale and violent star formation, resulting in massive blow-outs
that drag dark matter out due to gravitational interactions, thus
destroying the primordial CDM cusp.  The exact physics of feedback are
not understood, and the models usually resorts to an empirical
description that is fine-tuned to e.g.\ fit the Tully-Fisher relation
or other observational constraints.  

Nevertheless, let us explore the implications of feedback: using the
amount of energy produced by supernovae and making some standard
assumptions one can work out an expression for the amount of matter
expelled by the effects of supernovae (see e.g.~Eq.~(A4) in
\citealt{frankvdb}). Feedback then requires the baryonic mass to drag along
a similar amount of dark matter (modulo some efficiency factor that is
unlikely to be much larger than unity).

Assuming CDM halos, we find for $V_{200}\simeq 55$ \kms\ that $M_{200} = 5.2
\times 10^{10}\ M_{\odot}$. Equation A4 in \citet{frankvdb} then yields that
the expelled baryonic mass is $3.6 \times 10^9\ M_{\odot}$. Assuming
that the efficiency factor is unity (every unit mass of baryons drags
along one unit mass of dark matter) we find that the amount of dark
matter relocated is about equal to the amount of dark matter currently
observed within the outermost radius of the observed disk. We are thus
dealing with a major structural re-organisation of a galaxy.

Even if we assume the maximum possible maximum disk \MLstar=0.95 we
still only find an observed baryonic mass of $5.4\times 10^8\
M_{\odot}$ (though not physically motivated, this is a hard upper
limit on the observed baryonic mass).  Assuming an efficiency factor
of unity, it seems NGC 6822 must have expelled over $\sim 6$ times its
currently observed baryonic mass.
 
There has been some discussion in the past whether or not stellar
feedback can remove a significant fraction of the mass of a galaxy to
the Intergalactic Medium. \citet{gnedin}, as well as \citet{maclow}
showed that this process is unable to explain the observed cores, and
is also inconsistent with other observational constraints.
According to numerical simulation of stellar feedback by
\citet{maclow} the potential mass loss is a function of the total
galaxy mass; for galaxies with gas masses $< 10^6\ M_\odot$ a galaxy
can potentially destroy iself (`blow-away' scenario) whereas for gas
masses between $10^6$ and $10^9\ M_{\odot}$ some material may be
removed (`blow-out', see their Fig. 1). According to these models we
do not expect signifcant mass loss in NGC 6822 ($M_{gas}=1.5 \cdot
10^8$) due to violent star formation.

From an observer's perspective, there is relatively little evidence
that mass loss does indeed occur in dwarf galaxies. Although strong
outflow signatures are observed in some cases in dwarf galaxies there
are only few cases where outflow material may permanently escape the
galactic gravitational potential (e.g.\ \citealt{martin98}).

A further problem with expelling a lot of mass are the global dynamics
of the galaxy: if one were to have a huge mass loss of order a few
times the total galaxy mass, one would certainly expect to find some
clear evidence in the velocity field. The blown-out gas is such a case
unlikely to end up as an undisturbed regularly rotating \HI disk.
Furthermore, the stellar record shows no evidence for increases in the
SFR with large factors at any time during the past $\sim 9$ Gyr
\citep{gallart2}. It thus seems somewhat surprising that if feedback
had occurred, it would have managed to hide all traces of a violent
past.

In summary, we conclude that potential feedback of several times the
total baryonic mass of NGC\,6822 is unlikely.  The observed rotation
curve of NGC 6822, combined with the $K$-band and \HI data thus lead
to a picture inconsistent with a combination of cuspy halo and
feedback.

\section{Conclusions}

We obtained high resolution rotation curves for the local group dwarf
irregular galaxy NGC 6822. By fitting both pseudo-isothermal and NFW
halo models, it was found that the pseudo-isothermal model fits the
data successfully. NGC 6822 is not maximum disk and is dominated by
dark matter. The highest resolution rotation curve has some 250
independent points and beam-smearing effects are not an issue. The
position of the dynamical center is unambiguously determined. There is
no indication of a dark matter cusp down to scales of $\sim 20$ pc.

Trying to explain the observed curve with a combination of feedback
and CDM halos leads to several inconsistencies. Feedback and CDM seem
to imply that NGC 6822 has lost about 6 times the currently observed
amount of baryons. There is however no evidence for such violent
processes. We therefore conclude that (i) the dark matter distribution
in NGC 6822 is unlikely to have been affected by feedback, and (ii) is
best described by a model with a constant density core.

\section*{Acknowledgements} We thank Steve Schneider for providing us with the 2MASS 
images. We thank S\'ebastien Blais-Ouellette for constructive comments
that helped improve this paper.  This publication makes use of data
products from the Two Micron All Sky Survey, which is a joint project
of the University of Massachusetts and the Infrared Processing and
Analysis Center/California Institute of Technology, funded by the
National Aeronautics and Space Administration and the National Science
Foundation.

\clearpage

%\begin{figure*}
%\psfig{figure=b12.diffgaumom.ps,width=0.7\hsize}
%\caption{Histogram of the differences between the intensity-weighted and 
%gaussian velocity fields of the B12 data. The difference can be well described
%by a Gaussian with an average value of $-0.3$ \kms, and a RMS of $1.0$ \kms.
%The two velocity fields thus agree to well within one channelspacing.
%\label{gauhist}}
%\end{figure*}

\clearpage

%\begin{figure}
%\psfig{figure=b24.minunif.ps,width=0.5\hsize}
%\caption{Minimum disk fit for B24 assuming NFW model. Uniform weighting was used.\label{b24unif}}
%\end{figure}

%\begin{figure}
%\psfig{figure=slopehisto.ps,width=0.5\hsize}
%\caption{Histogram of inner slopes of mass density profile of NGC 6822
%\label{slopehisto}}
%\end{figure}

\clearpage
\begin{table*}
\begin{minipage}{105mm}
\caption{Data cube properties \label{datasets}}
\begin{tabular}{@{}lccccc}
\hline
 & B96 & B48 & B24 & B12 & B08 \\
\hline
beam size $('')$& 349.4 $\times$ 96   & 174.7 $\times$ 48  & 86.4 $\times$ 24  & 42.4 $\times$ 12  & 28.3 $\times$ 8   \\
pixel size $('')$&32 $\times$ 32    &16 $\times$ 16   & 8 $\times$ 8  & 4 $\times$ 4  & 2.5 $\times$ 2.5      \\
%$1\sigma$ noise channel$^{-1}$&  x & x  & x  & x  & x  \\
%(mJy/beam) &   &   &   &   &   \\
\hline
\end{tabular}

All data cubes presented here have a channel separation 
of 1.6 \kms, and an effective velocity resolution of 1.9 \kms.
\end{minipage}
\end{table*}

\begin{table*}
\begin{minipage}{160mm}
\caption{Adopted ROTCUR parameters \label{fits}}
\begin{tabular}{@{}llllll}
\hline
 & B96 & B48 & B24 & B12 & B08 \\
\hline
\XPOS  &$19^h44^m57.54^s$&$19^h44^m58.76^s$&$19^h44^m58.71^s$ &$19^h44^m58.04^s$& $19^h44^m58.91^s$\\
\YPOS  &$-14\degree 49^m13.0^s$ & $-14\degree 49^m17.6^s$ &$-14\degree 49^m21.7^s$ &$-14\degree 49^m18.9^s$ & $-14\degree 49^m23.8^s$ \\
\VSYS & $-53.3$ \kms & $-54.7$ \kms & $-54.7$  \kms & $-54.4$  \kms & $-54.6$  \kms\\
\PA  & {\it 0-1248:} 109\degree        & {\it 0-480:} 127-105\degree &  {\it 0-480:} 130-105\degree & {\it 0-456:} 140-103\degree &{\it 0-480:} 150-106\degree \\
      & {\it 1344-2592:} 109-138\degree & {\it 480-2400:} 105-130\degree & {\it 480-2136:} 105-131\degree &  {\it 456-1992:} 103-130\degree& {\it 480-680:} 106\degree\\
 & &  &  &  &{\it 688-1992:} 106-130\degree \\
\INCL  &   58.7\degree      & 59.9\degree & 58.7\degree & {\it 0-1056:} 58\degree & {\it 0-1096:} 57\degree \\ 
      &                    &             &    & {\it 1056-1992:} 58-68\degree & {\it 1104-1992:} 57-68\degree\\
\hline
\end{tabular}

Italics indicate a range in radius in arcseconds. The pair of numbers following it indicate the linear range in parameter over this radial range.
\end{minipage}
\end{table*}

\begin{table*}
\begin{minipage}{90mm}
\caption{NFW Rotation curve fits \label{nfwfits}}
\begin{tabular}{@{}llrrrrrr}
\hline
{Resolution} &  {model} &  {$c$} &  {$\Delta c$} &  {$V_{200}$} &  {$\Delta V_{200}$} &  {\MLstar} & {$\chi^2_{red}$} \\
\hline
B96 		& min & 3.7 & 1.2    &130.2 & 41.8   &  0  &  2.73 \\
		& mpg & 2.8 & 1.4    &157.0 & 76.6   &  0  &  2.49 \\
    		& con & 0.1 & ...    & 810.3& 14.1   &0.35 &  2.89 \\
    		& max & 0.1 & ...    & 721.9& 20.3   &0.95 &  5.97 \\
B96 inner 	& min & 0.1 & ...    & 878.0& 61.6   & 0   &  3.81 \\
		& mpg & 0.1 & ...    &1791.7& 56.7   & 0   &  3.24 \\
		& con & 0.1 & ...    & 636.9& 57.8   &0.35 &  3.36 \\
		& max & 0.1 & ...    & 377.5& 65.7   &0.95 &  4.36 \\
B48 		& min & 2.1 & 1.3    & 217.9&123.4   &  0  &  3.15 \\
		& mpg & 1.8 & 1.3    & 225.4&141.0   &  0  &  2.56 \\
    		& con & 0.1 & ...    & 730.6& 14.1   &0.35 &  5.91 \\
    		& max & 0.1 & ...    & 696.1& 16.9   &0.48 &  8.50 \\
B48 inner 	& min & 0.1 & ...    & 899.6& 26.4   & 0   &  5.99 \\
		& mpg & 0.1 & ...    & 826.8& 23.6   & 0   &  4.82 \\
		& con & 0.1 & ...    & 633.6& 33.0   &0.35 &  9.38 \\
		& max & 0.1 & ...    & 563.2& 37.1   &0.48 & 11.84 \\ 
B24 		& min & 0.1 & ...    & 879.5& 11.0   &  0  &  3.58 \\
		& mpg & 0.1 & ...    & 814.0& 10.0   &  0  &  2.99 \\
    		& con & 0.1 & ...    & 716.7& 15.3   &0.35 &  6.95 \\
    		& max & 0.1 & ...    & 698.5& 16.7   &0.42 &  8.27 \\
B24 inner 	& min & 0.1 & ...    & 876.5& 27.2   & 0   &  7.95 \\
		& mpg & 0.1 & ...    & 795.1& 24.7   & 0   &  6.56 \\
		& con & 0.1 & ...    & 591.3& 29.4   &0.35 &  9.31 \\
		& max & 0.1 & ...    & 551.1& 30.5   &0.42 & 10.02 \\
B12 		& min & 2.3 & 1.1    & 193.1& 92.2   &  0  &  5.74 \\
		& mpg & 1.6 & 1.4    & 237.1& 174.3  &  0  &  4.42 \\
    		& con & 0.1 & ...    & 630.8& 11.5   &0.35 &  6.15 \\
    		& max & 0.1 & ...    & 630.8& 11.5   &0.35 &  6.15 \\
B12 inner 	& min & 0.1 & ...    & 889.3& 19.2   & 0   & 10.61 \\
		& mpg & 0.1 & ...    & 792.0& 17.4   & 0   &  8.69 \\
		& con & 0.1 & ...    & 592.4& 20.1   &0.35 & 11.65 \\
		& max & 0.1 & ...    & 592.4& 20.1   &0.35 & 11.65 \\ 
B08 		& min & 0.1 & ...    & 879.3& 10.2   &  0  &  8.13 \\
		& mpg & 0.1 & ...    & 797.0&  9.3   &  0  &  6.73 \\
    		& con & 0.1 & ...    & 674.5& 11.7   &  0  & 10.69 \\
    		& max & 0.1 & ...    & 732.6& 10.3   &0.18 &  8.24 \\
B08 inner 	& min & 0.1 & ...    & 898.7& 20.6   & 0   & 15.52 \\
		& mpg & 0.1 & ...    & 792.7& 19.0   & 0   & 13.28 \\
		& con & 0.1 & ...    & 590.7& 22.3   &0.35 & 18.11 \\
		& max & 0.1 & ...    & 688.2& 20.6   &0.18 & 15.52 \\ 
\hline
\end{tabular}

{``min'': minimum disk; ``mpg'': minimum disk+gas; ``con'': constant \MLstar; ``max'': maximum disk. Dots (...) indicate the parameter was fixed during fitting.}
\end{minipage}
\end{table*}

\begin{table*}
\begin{minipage}{90mm}
\caption{ISO Rotation curve fits \label{isofits}}
\begin{tabular}{@{}llrrrrrr}
\hline
 {Resolution} &  {model} &  {$R_C$} &  {$\Delta R_C$} &  {$\rho_0$} &  {$\Delta \rho_0$} &  {\MLstar} & {$\chi^2_{red}$} \\
\hline

B96 		& min & 1.63 & 0.03  & 42.4 &  1.1   &  0  &  0.18 \\
		& mpg & 1.71 & 0.04  & 37.1 &  1.1   &  0  &  0.24 \\
    		& con & 2.12 & 0.06  & 27.1 &  0.9   &0.35 &  0.28 \\
    		& max & 3.10 & 0.22  & 16.7 &  1.0   &0.95 &  1.08 \\
B96 inner 	& min & 1.50 & 0.05  & 45.0 &  1.2   & 0   &  0.05 \\
		& mpg & 1.47 & 0.05  & 41.3 &  1.2   & 0   &  0.05 \\
		& con & 2.10 & 0.17  & 26.4 &  1.2   &0.35 &  0.10 \\
		& max &$\infty$&...  & 11.2 & 0.7    &0.95 &  0.55 \\
B48 		& min & 1.40 & 0.03  & 47.9 &  1.2   &  0  &  0.80 \\
		& mpg & 1.40 & 0.03  & 44.5 &  1.3   &  0  &  1.01 \\
    		& con & 2.03 & 0.05  & 26.2 &  0.7   &0.35 &  0.67 \\
    		& max & 2.37 & 0.07  & 21.7 &  0.6   &0.48 &  0.87 \\
B48 inner 	& min & 1.22 & 0.06  & 53.5 &  2.6   & 0   &  1.19 \\
		& mpg & 1.12 & 0.06  & 53.3 &  2.9   & 0   &  1.17 \\
		& con & 1.89 & 0.21  & 27.2 &  2.0   &0.35 &  1.55 \\
		& max & 2.70 & 0.55  & 20.3 &  1.7   &0.48 &  1.95 \\ 
B24 		& min & 1.51 & 0.03  & 44.6 &  1.0   &  0  &  0.99 \\
		& mpg & 1.58 & 0.04  & 39.5 &  1.0   &  0  &  1.31 \\
    		& con & 2.32 & 0.06  & 23.4 &  0.6   &0.35 &  1.00 \\
    		& max & 2.53 & 0.07  & 21.1 &  0.6   &0.42 &  1.04 \\
B24 inner 	& min & 1.31 & 0.07  & 49.8 &  2.1   & 0   &  1.06 \\
		& mpg & 1.20 & 0.06  & 48.7 &  2.2   & 0   &  0.96 \\
		& con & 2.27 & 0.34  & 24.1 &  1.7   &0.35 &  1.60 \\
		& max & 2.88 & 0.65  & 20.6 &  1.6   &0.42 &  1.84 \\
B12 		& min & 1.24 & 0.02  & 53.2 &  0.8   &  0  &  0.78 \\
		& mpg & 1.23 & 0.02  & 48.3 &  0.9   &  0  &  0.92 \\
    		& con & 1.84 & 0.04  & 26.6 &  0.6   &0.35 &  1.21 \\
    		& max & 1.84 & 0.04  & 26.6 &  0.6   &0.35 &  1.21 \\
B12 inner 	& min & 1.19 & 0.04  & 55.0 &  1.7   & 0   &  1.53 \\
		& mpg & 1.06 & 0.04  & 54.4 &  1.8   & 0   &  1.35 \\
		& con & 1.78 & 0.15  & 27.2 &  1.5   &0.35 &  2.50 \\
		& max & 1.78 & 0.15  & 27.2 &  1.5   &0.35 &  2.50 \\
B08 		& min & 1.29 & 0.02  & 52.4 &  0.8   &  0  &  1.11 \\
		& mpg & 1.37 & 0.02  & 44.6 &  0.7   &  0  &  1.00 \\
    		& con & 2.01 & 0.05  & 25.8 &  0.6   &0.35 &  1.84 \\
    		& max & 1.66 & 0.03  & 33.7 &  0.6   &0.18 &  1.27 \\
B08 inner 	& min & 1.43 & 0.05  & 49.0 &  1.2   & 0   &  1.81 \\
		& mpg & 1.39 & 0.05  & 44.3 &  1.3   & 0   &  1.73 \\
		& con & 3.01 & 0.40  & 22.2 &  1.0   &0.35 &  3.06 \\
		& max & 1.88 & 0.11  & 31.5 &  1.1   &0.18 &  2.21 \\ 
\hline
\end{tabular}

{``min'': minimum disk; ``mpg'': minimum disk+gas; ``con'': constant \MLstar; ``max'': maximum disk.}
\end{minipage}
\end{table*}

%\begin{deluxetable}{llrrrrrr}
%\tablecaption{NFW minimum disk Rotation curve fits, uniform weighting\label{uniffits}}
%\tablewidth{0pt}
%\tablehead{
% {Resolution} &  {model} &  {$c$} &  {$\Delta c$} &  {$V_{200}$} &  {$\Delta V_{200}$} &  {\MLstar} & {$\chi^2_{red}$} 
%}
%\startdata
%B96 		& min & 4.4 & 1.0    & 103.5 & 21.6   &  0  & 9.7 \\
% 		& mpg & 3.8 & 1.0    & 110.9 & 28.1   &  0  & 8.6 \\
% 		& con & 1.2 & 3.0    & 283.9 & 511.6  & 0.35& 9.9 \\
%B48 		& min & 5.7 & 0.7    &  77.2 &  7.5   &  0  & 9.4\\
% 		& mpg & 5.7 & 0.7    &  72.7 &  7.0   &  0  & 8.6\\
% 		& con & 3.3 & 1.0    & 109.5 & 31.2   & 0.35&11.6\\
%B24 		& min & 0.0 & 0.1    &1010.3 & 21.0   &  0  & 5.2 \\
% 		& mpg &-0.3 & 42.0   &1587.3 &$\infty$&  0  & 4.3 \\
% 		& con &-0.4 &182.9   &1643.2 &$\infty$& 0.35& 9.3 \\
%B12 		& min & 3.6 & 0.6    & 119.0 & 18.9   &  0  & 6.8 \\
% 		& mpg & 2.5 & 0.7    & 156.0 & 41.4   &  0  & 5.0 \\
% 		& con &-0.4 &82.7    &1422.3 &$\infty$& 0.35& 7.5 \\
%B08 		& min & 3.6 & 0.6    & 123.4 & 19.8   &  0  & 9.7\\
% 		& mpg & 1.5 & 1.2    & 255.9 &165.3   &  0  & 7.8\\
% 		& con &-0.4 &73.5    &1483.3 &$\infty$& 0.35&10.8\\
%\enddata
%\end{deluxetable}

\begin{table}
\caption{Inner mass density slopes \label{slope}}
\begin{tabular}{@{}lrr}
\hline
 {Resolution} &  {$\alpha$} &  {$\Delta \alpha$} \\
\hline
B96&$-0.13$&	0.14\\
B48&$+0.02$&	0.12\\
B24&$+0.22$&	0.35\\
B12&$-0.04$&	0.09\\
B08&--&		--	\\
\hline
\end{tabular}
\end{table}
\end{document}